\documentclass[apj]{emulateapj}
\usepackage{multirow}
\usepackage{longtable}
\usepackage{rotating}

\def\gtorder{\mathrel{\raise.3ex\hbox{$>$}\mkern-14mu
             \lower0.6ex\hbox{$\sim$}}}
\def\ltorder{\mathrel{\raise.3ex\hbox{$<$}\mkern-14mu
             \lower0.6ex\hbox{$\sim$}}}

\slugcomment{Draft of \today}

\shorttitle{Fresh and Weathered Asteroid Pairs}

\shortauthors{Polishook et al.}

\begin{document}

\title{Observations of ``Fresh" and Weathered Surfaces on Asteroid Pairs and Their Implications on the Rotational-Fission Mechanism}
\author{D.~Polishook\altaffilmark{1},
N. Moskovitz \altaffilmark{1},
R.~P. Binzel \altaffilmark{1},
F.~E. DeMeo \altaffilmark{1,2},
D. Vokrouhlick{\'y} \altaffilmark{3},
J. {\v{Z}}i{\v{z}}ka \altaffilmark{3},
D. Oszkiewicz \altaffilmark{4},
}

\altaffiltext{1}{Department of Earth, Atmospheric, and Planetary Sciences, Massachusetts Institute of Technology, Cambridge, MA 02139, USA}
\altaffiltext{2}{Harvard-Smithsonian Center for Astrophysics, 60 Garden Street, Cambridge, MA 02138, USA}
\altaffiltext{3}{Institute of Astronomy, Charles University, Prague, V Hole{\v{s}}ovi{\v{c}}k{\'a}ch 8, CZ - 18000 Prague 8, Czech Republic}
\altaffiltext{4}{Astronomical Observatory Institute, Faculty of Physics, A. Mickiewicz University, S{\l}oneczna 36, 60-286, Pozna{\'n}, Poland}

\begin{abstract}
	The rotational-fission of a ``rubble-pile" structured asteroid can result in an ``asteroid pair" - two un-bound asteroids sharing nearly identical heliocentric orbits. Models suggest that this mechanism exposes material from below the progenitor surface that previously had never have been exposed to the weathering conditions of space. Therefore, the surfaces of asteroid pairs offer the opportunity to observe non-weathered ``fresh" spectra.

	Here we report near-infrared spectroscopic observations of 31 asteroids in pairs. In order to search for spectral indications of fresh surfaces we analyze their spectral slopes, parameters of their $1 \mu m$ absorption band and taxonomic classification. Additionally, through backward dynamical integration we estimate the time elapsed since the disintegration of the pairs’ progenitors.

	Analyzing the 19 ordinary chondrite-like (S-complex) objects in our sample, we find two Q-type asteroids (19289 and 54827) that are the first of their kind to be observed in the main-belt of asteroids over the full visible and near-infrared range. This solidly demonstrates that the Q-type taxonomy is not limited to the NEA population.

	The pairs in our sample present a range of fresh and weathered surfaces with no clear evidence for a correlation with the ages of the pairs. However, our sample includes ``old" pairs ($2x10^6 \ge$ age $\ge 1x10^6$ years) that present relatively low, meteoritic-like spectral slopes ($<0.2\%$ per $1 \mu m$). This illustrates a timescale of at least $\sim 2$ million years before an object develops high spectral slope that is typical for S-type asteroids.

	We discuss three mechanisms that explain the existence of weathered pairs with young dynamical ages and find that the ``secondary fission" model (Jacobson \& Scheeres 2011) is the most robust with our observations. In this mechanism an additional and subsequent fission of the secondary component contributes the lion share of fresh material that re-settles on the primary's surface and recoats it with fresh material. If the secondary breaks loose from the vicinity of the primary before its ``secondary fission", this main source of fresh dust is avoided. We prefer this secondary fission model since {\it i)} the secondary members in our sample present ``fresh" parameters that tend to be ``fresher" than their weathered primaries; {\it ii)} most of the fresh pairs in our sample have low size ratios between the secondary and the primary; {\it iii)} 33\% of the primaries in our sample are fresh, similar to the prediction set by the secondary fission model (Jacobson \& Scheeres 2011); {\it iv)} known satellites orbit two of the pairs in our sample with low size ratio (D2/D1) and fresh surface; {\it v)} there is no correlation between the weathering state and the primary shape as predicted by other models.


\end{abstract}

\keywords{
Asteroids; Asteroids, rotation; Asteroids, surfaces; Rotational dynamics; Spectroscopy}

\section{Introduction and Motivation}
\label{sec:Introduction}

\subsection{Dynamics and Formation of Asteroid Pairs}
\label{sec:pairsFormation}

	Pairs of asteroids move about the Sun on very similar orbits (Vokrouhlick{\'y} \& Nesvorn{\'y} 2008), but, unlike binary asteroids\footnote{Binary asteroids are two objects revolve about a common center of mass, which itself moves about the Sun (e.g., Merline et al. 2002a, Richardson \& Walsh 2006, Pravec et al. 2006, Taylor \& Margot 2011).}, are gravitationally unbound. The orbits of paired asteroids are so similar that they cannot be a mere coincidence (Vokrouhlick{\'y} \& Nesvorn{\'y} 2008, 2009, Pravec \& Vokrouhlick{\'y} 2009). Moreover, using backwards orbital integrations have shown that members of each pair were in the same location in space sometime within the past few million years. This suggests a common origin for the components of each pair. Indeed, spectroscopic observations and broadband photometry studies have shown that members of observed pairs have similar spectra or colors (Moskovitz 2012, Duddy et al. 2012, 2013). It was also found that asteroid pairs are not correlated to a specific type of composition or taxonomic class (Moskovitz 2012).

	Pair formation by collision has been rejected due to the low relative velocity between components at the time of their formation (e.g., Vokrouhlick{\'y} \& Nesvorn{\'y} 2008, 2009, Pravec \& Vokrouhlick{\'y} 2009). Rather, this low velocity supports a model of a gentle separation of an unstable binary asteroid configuration. This is further supported by the distribution of the mass ratio between the members of each pair that is complementary to the distribution of gravitationally bound binary asteroids (Vokrouhlick{\'y} \& Nesvorn{\'y} 2008). Modeling suggests that pairs form by the fission of a fast-rotating aggregate-like asteroid (with the so-called ``rubble-pile" structure) into two objects (e.g., Scheeres 2007, 2009, Jacobson \& Scheeres 2011). Finally, photometric measurements (Pravec et al. 2010) showed that rotation periods of the larger members of asteroid pairs are correlated with the mass ratio in a way that matches the rotational-fission mechanism: (i) if the secondary (the smaller member) is massive enough, it carries a significant amount of angular momentum and the rotation rate of the primary (the larger member) will decelerate; (ii) if the secondary is not massive, the primary will continue to rotate fast. Furthermore, these measurements also confirmed that there is a limit to the secondary mass fraction at $\sim20\%$ of the primary, as previously predicted by theoretical models. Larger secondaries do not have sufficient energy to leave the primary; thus they remain in its vicinity, forming binary asteroids (Pravec et al. 2010, Scheeres 2007).

\subsection{The Rotational-Fission Mechanism}
\label{sec:rotationalFission}

	The main process to accelerate asteroids' spins is the Yarkovsky-O'Keefe-Radzievsky-Paddack effect, also known as the YORP effect (Rubincam 2000, Bottke et al. 2006). The YORP effect is a radiation torque imposed on a rotating body due to the asymmetric reflection and re-emission of sunlight. The relatively short evolution timescale of 1 to 10 Myr for small-sized asteroids (with diameter smaller than $\sim10$ km), confirmed by direct detections (e.g., Lowry et al. 2007, Taylor et al. 2007, Kaasalainen et al. 2007, {\v{D}}urech et al. 2008, 2012), makes the YORP effect a very efficient mechanism to control the spins of asteroids among the near-Earth asteroids (NEAs) and main-belt asteroids (MBAs; e.g., Pravec et al. 2008, Polishook \& Brosch 2009). While the rotation of an asteroid can also be spun-up by sub-catastrophic impacts, the YORP effect seems to be a more robust and efficient process for small-sized asteroids\footnote{We should note that theoretically the YORP effect can also spin-down asteroid spins, depending on their physical parameters; however, this scenario is irrelevant for the rotational-fission mechanism.} (Marzari et al. 2011).

	When the accelerated spin of the asteroid reaches the critical spin for a ``rubble-pile" object (at about 2.2 hours per rotation; Richardson et al. 1998, Pravec \& Harris, 2000), the asteroid fissions (Margot et al. 2002). However, different scenarios of the rotational fission process have been proposed. For instance, Walsh et al. (2008, 2012) present a model in which the fast rotation transports material towards the equator and gradually forms a near-equatorial ridge (as evidenced, e.g., by the diamond-shape of asteroid (66391) 1999 KW4; Ostro et al. 2007, Harris et al. 2009; and other objects). If continued, this process can eject part of the equatorial mass, where it can re-accumulate into a satellite. Using this model, Walsh et al. were able to theoretically produce satellites and diamond-shape objects as seen in nature by observations. However, it is unclear if the ejected material has enough time in orbit around the asteroid to be accumulated into a satellite. In addition, Holsapple (2010) using granular theory finds that mass loss should not occur at the equator but rather the shape of the body would deform until interior failure occurs. Furthermore, the elongated shapes of some asteroid pairs (Pravec et al. 2010) do not match the diamond shapes resulted by Walsh et al. model.

	Alternatively, Scheeres (2007, 2009) describes a model of a coarser internal structure of the parent body that consists of a set of larger components. His model suggests that the rotational-fission mechanism can result in the loss of a significant part of the fast-rotating body so that the ejected component (the secondary member) will start its own course around the Sun. Jacobson \& Scheeres (2011) further developed this model and suggested that the secondary itself might disintegrate since it is under the pressure of the primary's tidal forces during the tens of days after its detachment and before it is lost in space. A fission of the secondary might form a third body that can crash into the primary, fall back on the secondary, or be lost to space. As the third body leaves the system it carries with it the excess of angular momentum, by that stabilizing the orbit of the secondary object around the primary, allowing them to become gravitationally bound as a binary asteroid.

	The model of Walsh et al. and the model of Scheeres and Jacobson differs in duration over which the fission process takes place: the first is a gradual and slow process that can take one or more spin-up pulses induced by the YORP effect, stretching out over a long time interval (hundreds of ky to Mys). The fission by the second model is immediate, and a few days up to tens of days are needed before the ejected component is lost. This scenario is also more violent than the gradual model, since more energy is needed to remove a significant part of the asteroid, and this is probably followed with the removal of dust and debris that sink back on the main body and recoating it. Further disintegration of the secondary, and possible impacts between the ejected components to the primary object, probably results with even more dust and debris. The recent observation of the main belt object P/2013 P5 that presented a dusty structure of multi-tails and a coma (Jewitt et al. 2013) can be explained by a rotational-fission event of a fast rotating asteroid and the following fission of its secondary member, thus it supports the fast model.

	While more diamond-shaped, fast-rotating asteroids have been found in recent years, supporting the Walsh et al. model, the Scheeres' model helps to better explain the relatively large secondaries of asteroid pairs and the above mentioned strong correlation between the rotation period of the primary and the mass ratio of the two components. If the two models are valid, it is unknown what conditions will favor one mechanism over the other, and which is the more frequent scenario among asteroids. One way of probing the fission models is provided by spectral observations. This is because the extent of excavation and transportation of material following rotational-fission might be revealed on asteroids of the Ordinary Chondrite (OC) type (part of the so-called S-complex in the jargon of asteroid taxonomy) by identifying how much age-related alterations of their spectra are seen. If properly understood, such data might in principle help distinguish between the different rotational-fission models.

\subsection{Space Weathering}
\label{sec:spaceWeathering}

	According to the previously mentioned models, the fission process takes its toll on the asteroid - boulders and rocks are shifted, regolith and dust are disturbed. During this ``gardening" process, sub-surface layers might have been excavated and exposed to space. In the case of OC asteroids, the exposed material might have fresh spectral properties that were not modified by the ``space weathering" mechanism. This mechanism, caused by solar wind, cosmic rays and micrometeorite bombardment, alters the top layer on atmospheres-less planetary surfaces, causing them to display a ``weathered", darker and redder reflectance spectrum (e.g., Clark et al. 2002). Brunetto et al. (2006) found that an exponential curve best mimics the reddening effect of the space weathering mechanism:

\begin{equation}
W({\lambda}) = F({\lambda})exp(C_{S}/{\lambda}),
\label{eq:BrunettoEq}
\end{equation}

where F is the fresh reflectance per wavelength $\lambda$, W is the weathered reflectance and the power-law $C_{S}$ is the extent of the space weathering. Applying this empirical rule on a fresh meteoritic spectrum results in the increase of the spectral slope and decrease of the depth of the $1 \mu m$ absorption band (although it does not decrease the surface albedo).

	Non-weathered OC minerals were brought to Earth by the Hayabusa spacecraft from the asteroid (25143) Itokawa (Noguchi et al. 2011), even though the asteroid presents a weathered spectrum (Binzel et al. 2001). This gave the ultimate observational support for the modifications of the top layer of asteroids by the space weathering mechanism and concluded a long debate in the community (Chapman 2004).

	OC asteroids with non-weathered surfaces have been observed among NEAs and were dubbed as Q-type asteroids (e.g., Tholen 1984, Binzel et al. 1996), while Sq-type asteroids are in an intermediate state (DeMeo et al. 2009). A couple of Q-types have been found by spectral observations in the visible range among the small members in the Datura dynamical cluster (e.g., Moth{\'e}-Diniz \& Nesvorn{\'y} 2008), a young family of asteroids formed by a collision $\sim450\pm50$ kyr ago (e.g., Nesvorn{\'y} et al. 2006, Vokrouhlick{\'y} et al. 2009). Using visible-wavelength broadband photometry Rivkin et al. (2011) found some asteroids with Q-type colors among asteroids in the much older and larger Koronis family that was also formed by a collision. Thomas et al. (2011) have found a trend in spectral slope for objects 1-5 km that shows the transition from Q- to S-type among the Koronis family as well. Binzel et al. (2010) and Nesvorn{\'y} et al. (2010) suggested that tidal forces from the terrestrial planets could expose fresh areas when the asteroid has a close orbital intersection with the Earth or Venus and DeMeo et al. (2014) pointed out the possibility this mechanism might be valid for Mars. However, we do not know which mechanism of forming Q-type asteroid surfaces is the most efficient, planetary encounters or collisions, and what is the role of rotational-fission in this context.

	All these studies have established a link between exposure of fresh material to violent processes (such as collisions, planetary tidal forces, etc.), but some arrived at a different determination of the timescale of the space weathering mechanism (see below). Furthermore, it seems that different types of weathering mechanisms exist. For example, while the regolith on the moon becomes redder and darker with weathering, some asteroids, such as (243) Ida (that was studied by the space mission Galileo), becomes only redder with no albedo modifications, and others, such as (433) Eros, becomes only darker with no color alteration (e.g., Gaffey 2010). This suggests that different types of OC reacts differently to space weathering and one should not generalize all its dependencies and effects.

\subsection{Space Weathering Timescale}
\label{sec:spaceWeatheringTimescale}

	Current estimations of the timescale of space weathering differ dramatically from one another, and range between 50 ky (e.g., Sasaki et al. 2001) to $10^9$ years (e.g., Willman et al. 2010). This wide range probably includes different stages of weathering, performed by different agents (Solar wind, micrometeorite bombardment, etc.). For example, Vernazza et al. (2009) suggested that the Solar wind is the origin of the rapid reddening of asteroid surfaces, compared to a slower effect caused by the micrometeorites. By measuring the spectral slopes of asteroids from young collisional families, Vernazza et al. (2009) suggested a maximal limitation of $\sim10^6$ years for the faster space weathering timescale. Nesvorn{\'y} et al. (2010) studied the orbital distributions of Q-type NEAs and their close approaches to the terrestrial planets and found a minimal timescale of $10^5$ years. These values make the asteroid pairs relevant for measuring the space weathering timescale because their ages (since the fission of their progenitors and the supposed exposure of fresh material) range from a few times $10^4$ years to a few times $10^6$ years. In this study we observed a sample of 31 asteroids in pairs, measured their infrared spectra, chose those of the OC type, and analyzed their spectral parameters in order to reveal their history and to search for a possible link with the processes of space weathering and rotational-fission.

\section{Observations and Reduction}
\label{sec:observations}

\subsection{Infrared Spectroscopy}
\label{sec:infraredSpectroscopy}

	We conducted a near infrared (0.8 to 2.5 $\mu m$) spectroscopic campaign for a sample of 31 asteroids in pairs. The candidate pair asteroids were taken from Pravec and Vokrouhlick{\'y} (2009), Vokrouhlick{\'y} (2009), Ro{\.{z}}ek et al. (2011) and Pravec (private communication\footnote{Petr Pravec (private communication) identified the asteroid pairs involving the primaries 8306, 4905, 16815 and 74096 with the method of Pravec and Vokrouhlick{\'y} (2009), updated with the use of mean elements from AstDyS-2.}). Most observations took place using SpeX, an imager and spectrograph mounted on the 3-m telescope of NASA's InfraRed Telescope Facility (IRTF; Rayner et al. 2003). Additional measurements of two asteroids were obtained with the 6.5m Magellan/Baade telescope of Las Campanas Observatory using FIRE, an equivalent spectrograph (Simcoe et al. 2013). A long slit with a 0.8 arcsec width was used and the objects were shifted along it in a A-B-B-A sequence to allow the measurement of the background noise. Observations were limited to low air mass values to reduce chromatic refraction that can change the spectral slope. The observational details are listed in Table~\ref{tab:ObsCircum}. The reduction of the raw SpeX images follows the procedures outlined in Binzel et al. (2010) and DeMeo et al. (2009). This includes flat field correction, sky subtraction, manual aperture selection, background and trace determination, removal of outliers, and a wavelength calibration using arc images. A telluric correction routine was used to model and remove telluric lines. Each spectrum was divided by a standard solar analog to derive the relative reflectance of the asteroid (stars are listed in Table~\ref{tab:ObsCircum}). Eight asteroids that were observed multiple times were additionally divided by a second star, of a G2 to G5 type, that were observed circa the time and coordinates of the asteroid. This additional normalization was used to erase slope differences due to atmospheric instability. Each normalized spectrum was then used to calculate the mean value of the reflectance.

\subsection{Visible Regime}
\label{sec:visibleRegime}

	While the reflectance in the near-infrared regime is essential for taxonomic and mineralogical analysis, the reflectance at visible wavelength (0.4 to 0.9 $\mu m$) is of no less importance. Since the absorption band at one $\mu m$, which is the main classification feature for asteroids of the S-complex group, stretches from approximately 0.7 to 1.3 $\mu m$, the knowledge of the band depth is lost if only the infrared regime is observed. While there are other band parameters that distinguish between S-type and Q-type spectra (such as the band center and width; see below), the band depth is a very important parameter to determine their exact classification.

	For 22 of the asteroid pairs, visible spectra were obtained from different sources:

{\it i)}	Spectral measurements for five asteroids were obtained with Magellan's 6.5m Clay telescope with the LDSS3 instrument, the 3m Lick telescope, and the 2.5m Nordic Optical Telescope (NOT). Observational details appear in Table~\ref{tab:ObsCircum}. The reduction processes were similar to those used for the IR data.

{\it ii)}	Broadband photometric colors using the SDSS griz filter set were obtained for two asteroids at Kitt-Peak's 2.1m telescope. Cousins/Johnson BVRI colors for three additional asteroids were observed with Wise Observatory's 0.46m (Brosch et al. 2008). Reduction was done in a standard way (details at Polishook \& Brosch 2008). To calibrate the photometry to a standard magnitude level we used Landolt stars (Landolt 1992) or local reference stars that appear in the SDSS catalog. For additional nine objects we used data published by Moskovitz (2012) who measured Cousins/Johnson BVRI colors and summarized SDSS's ugriz measurements (Juri{\'{c}} et al. 2007). We fit these values to the normalized reflectance scale by subtracting the solar brightness at a specific filter (values in Table~\ref{tab:SolarMagCalib}) from the relevant magnitude of the asteroid and translate the results into flux units. The derived values were normalized by the value at 0.55 $\mu m$ to give the normalized reflectance.

{\it iii)}	The visible spectra of three asteroids were measured and published by Duddy et al. (2013). Since we did not have direct access to this data, we used the spectrum of the best taxonomic fit that was found by the authors. The infrared spectra were stitched to the visible counterpart using the overlapping values between 0.8 to 0.95 $\mu m$. Because unity is set to 0.55 $\mu m$, the infrared reflectance was scaled to match the visible part.

\begin{figure}
\centerline{\includegraphics[width=8.5cm]{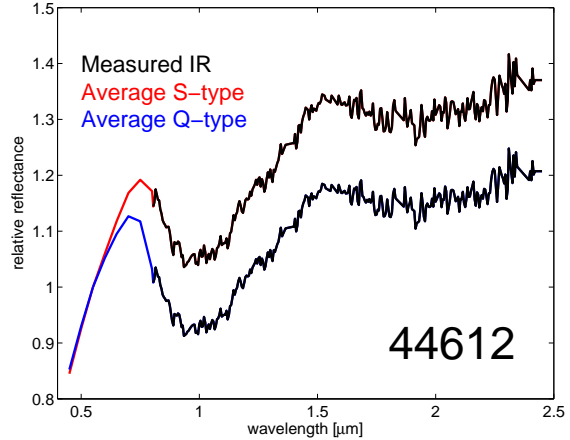}}
\caption{An example for stitching the visible spectrum (0.45 to 0.8 $\mu m$) of an average S-type and an average Q-type to the measured IR spectrum of 44612. The stitching was performed by scaling the near-IR spectrum to the visible spectrum, where the visible spectrum is normalized at 0.55 $\mu m$. Such a stitching was done for the six S-complex asteroids without any data in the visible regime.
\label{fig:Fig1}}
\end{figure}

	For six cases of OC spectra without visible reflectance we stitched two reflectance values of extreme cases: the visible reflectance of the S-type archetype, and the visible reflectance of the Q-type archetype, both from DeMeo et al. (2009). An example is shown in Fig.~\ref{fig:Fig1}. Any analysis work performed on these spectra was done separately on the two extremes and the uncertainty was adjusted to include both options.

\begin{deluxetable*}{ccccccccccc}
\tablecolumns{11}
\tablewidth{0pt}
\tablecaption{Observational details}
\tablehead{
\colhead{Name} &
\colhead{Date} &
\colhead{Exp} &
\colhead{Filter} &
\colhead{Telescope/CCD} &
\colhead{${\it R}$} &
\colhead{${\it \Delta}$} &
\colhead{${\it \alpha}$} &
\colhead{${\it V_{mag}}$} &
\colhead{Solar Analog} &
\colhead{Ref.} \\
\colhead{}          &
\colhead{}          &
\colhead{[min]}          &
\colhead{}    &
\colhead{}          &
\colhead{[AU]} &
\colhead{[AU]} &
\colhead{[deg]} &
\colhead{} &
\colhead{} &
\colhead{}
}
\startdata
1741 & 2013 03 07 & 16 & IR & IRTF/SpeX & 2.75 & 2.74 & 20.8 & 16.8 & L102-1081 & --- \\
1979 & 2013 03 06 & 24 & IR & IRTF/SpeX & 2.34 & 1.41 & 11.3 & 16.8 & L98-978 & --- \\
         & 2013 05 07 & 60 & Vis & NOT/ALFOSC & 2.41 & 2.09 & 24.7 & 18.2 & L102-1081 & --- \\
2110 & 2011 10 25 & 148 & IR & IRTF/SpeX & 1.96 & 0.97 & 3.9 & 15.0 & L93-101 & --- \\
         & 2011 10 26 & 128 & IR & IRTF/SpeX & 1.97 & 0.97 & 3.3 & 14.9 & L93-101 &  --- \\
         & 2010 02 28 & --- & BVRI & DuPont/SITe2k & 2.51 & 1.66 & 14.2 & 17.1 & --- & Moskovitz \\
2897 & 2013 03 07 & 16 & IR & IRTF/SpeX & 2.04 & 1.05 & 5.6 & 15.3 & L102-1081 & --- \\
3749 & 2012 01 22 & 58 & IR & IRTF/SpeX & 2.00 & 1.01 & 0.6 & 14.7 & L98-978 & --- \\
         & 2012 02 17 & 20 & Vis & Lick & 2.01 & 1.10 & 15.0 & 15.6 & L98-978 & --- \\
4765 & 2013 01 10 & 96 & IR & IRTF/SpeX & 1.84 & 1.53 & 32.4 & 17.3 & L105-56 & --- \\
         & 2013 01 11 & 88 & IR & IRTF/SpeX & 1.84 & 1.52 & 32.4 & 17.3 & L105-56 & --- \\
         & --- & --- & g'r'i'z' & SDSS & --- & --- & --- & --- & --- & Juri{\'{c}} et al. \\
4905 & 2013 08 08 & 16 & IR & IRTF/SpeX & 2.18 & 1.64 & 26.4 & 16.1 & L93-101 & --- \\
         &            &    &    &           &      &      &      &      & L110-361 & \\
         & 2013 09 12 &  & BVR & Wise/C18 & 2.17 & 1.30 & 17.7 & 15.3 & PG2331+055A & --- \\
5026 & 2012 04 21 & 20 & IR & IRTF/SpeX & 2.52 & 1.54 & 5.5 & 17.2 & L105-56 & --- \\
         & 2012 06 03 & 12 & Vis & Magellan/LSSD3 & 2.42 & 1.74 & 21.3 & 18.0 & SA105-56 & --- \\
6070 & 2012 05 26 & 18 & IR & IRTF/SpeX & 2.65 & 1.64 & 0.6 & 17.0 & L102-1081 & --- \\
         & 2013 10 03 & 40 & IR & IRTF/SpeX & 1.94 & 1.31 & 28.3 & 17.0 & Hya64 & --- \\
         & 2013 10 31 & 16 & IR & IRTF/SpeX & 1.98 & 1.12 & 19.0 & 16.4 & L98-978 & --- \\
         &            &    &    &           &      &      &      &      & L93-101 &  \\
         & 2012 06 01 & 12 & Vis & Magellan/LSSD3 & 2.64 & 1.63 & 2.7 & 17.2 & HD149182 & --- \\
8306 & 2013 09 07 & 16 & IR & IRTF/SpeX & 1.75 & 0.77 & 11.2 & 16.2 & L115-271 & --- \\
         &            &    &    &           &      &      &      &      & L110-361 & \\
         & 2013 09 12 &  & BVRI & Wise/C18 & 1.75 & 0.76 & 7.5 & 16.0 & PG2331+055A & --- \\
9068 & 2013 07 11 & 24 & IR & IRTF/SpeX & 1.62 & 1.54 & 37.4 & 17.0 & L113-271 & --- \\
         & 2013 09 12 &  & BVR & Wise/C18 & 1.56 & 1.09 & 40.0 & 16.2 & PG2331+055A & --- \\
10484 & 2011 10 27 & 140 & IR & IRTF/SpeX & 2.21 & 1.24 & 7.1 & 16.5 & L93-101 & --- \\
           &            &    &    &           &      &      &      &      & L113-276 & \\
           & 2013 05 07 & 60 & Vis & NOT/ALFOSC & 2.38 & 1.59 & 18.5 & 17.6 & L102-1081 & --- \\
13732 & 2014 01 07 & 50 & IR & IRTF/SpeX & 2.15 & 1.36 & 20.0 & 17.6 & Hya64 & --- \\
           &            &    &    &           &      &      &      &      & L98-978 & \\
           & 2011 02 24 & 52.5 & Vis & WHT & 2.49 & 2.09 & 23.4 & 19.1 & SA112 113 & Duddy et al. \\
15107 & 2013 01 17 & 48 & IR & IRTF/SpeX & 2.67 & 1.83 & 13.6 & 18.5 & L98-978 & --- \\
           & 2010 02 28 & --- & BVRI & DuPont/SITe2k & 2.59 & 1.77 & 14.9 & 18.4 & --- & Moskovitz \\
16815 & 2013 09 28 & 32 & IR & IRTF/SpeX & 2.62 & 1.65 & 7.4 & 16.3 & L112-1333 & --- \\
17198 & 2012 10 18 & 80 & IR & IRTF/SpeX & 2.38 & 1.39 & 0.8 & 17.6 & L93-101 & --- \\
           & 2012 10 19 & 48 & IR & IRTF/SpeX & 2.38 & 1.39 & 0.8 & 17.6 & L93-101 & --- \\
           & 2011 05 24 & 30 & Vis & WHT & 2.10 & 1.22 & 17.9 & 17.9 & SA112 113 & Duddy et al. \\
17288 & 2013 04 12 & 48 & IR & IRTF/SpeX & 2.65 & 1.68 & 6.2 & 17.8 & L98-978 & --- \\
           &            &    &    &           &      &      &      &      & 102-1081 & \\
           & --- & --- & g'r'i'z' & SDSS & --- & --- & --- & --- & --- & Juri{\'{c}} et al. \\
           & 2010 03 08 & --- & BVRI & Magellan/IMACS & 2.49 & 2.07 & 22.9 & 18.7 & --- & Moskovitz \\
19289 & 2012 09 11 & 36 & IR & IRTF/SpeX & 1.95 & 0.95 & 5.7 & 17.1 & L115-271 & --- \\
           & 2011 05 23 & 75 & Vis & WHT & 2.23 & 2.01 & 27.0 & 19.8 & SA112 113 & Duddy et al. \\
25884 & 2011 10 25 & 24 & IR & IRTF/SpeX & 1.93 & 0.96 & 8.7 & 16.5 & L93-101 & --- \\
           & 2011 10 26 & 56 & IR & IRTF/SpeX & 1.93 & 0.96 & 8.1 & 16.5 & Hya & --- \\
           & 2011 10 29 & 88 & IR & IRTF/SpeX & 1.94 & 0.96 & 6.1 & 16.4 & Hya & --- \\
38707 & 2013 05 12 & 48 & IR & IRTF/SpeX & 2.50 & 1.50 & 1.8 & 18.0 & L105-56 & --- \\
           &            &    &    &           &      &      &      &      & L107-684 & \\
(32957) & 2010 08 31 & --- & BVRI & DuPont/SITe2k & 2.33 & 2.05 & 25.6 & 20.6 & --- & Moskovitz \\
42946 & 2013 01 17 & 32 & IR & IRTF/SpeX & 2.61 & 1.80 & 14.9 & 17.8 & Hya64 & --- \\
           & 2013 02 16 &  & g'r'i'z' & KPNO2.1 & 2.58 & 2.11 & 21.4 & 18.3 & RU149F & --- \\
44612 & 2012 09 11 & 28 & IR & IRTF/SpeX & 1.82 & 0.81 & 2.0 & 16.6 & L115-271 & --- \\
52852 & 2012 12 17 & 34 & IR & IRTF/SpeX & 2.09 & 1.12 & 6.6 & 17.2 & Hya64 & --- \\
(250322) & --- & --- & g'r'i'z' & SDSS & --- & --- & --- & --- & --- & Juri{\'{c}} et al. \\
54041 & 2012 11 10 & 24 & IR & IRTF/SpeX & 2.27 & 1.30 & 7.8 & 17.4 & Hya64 & --- \\
           & 2012 12 14 & 48 & IR & IRTF/SpeX & 2.32 & 1.39 & 10.4 & 17.7 & Hya64 & --- \\
           & 2012 12 17 & 28 & IR & IRTF/SpeX & 2.32 & 1.41 & 11.7 & 17.8 & Hya64 & --- \\
           & 2012 12 19 & 46 & IR & IRTF/SpeX & 2.32 & 1.43 & 12.6 & 17.8 & Hya64 & --- \\
           & --- &  & g'r'i'z' & SDSS & --- & --- & --- & --- & --- & Juri{\'{c}} et al. \\
54827 & 2012 08 08 & 57 & IR & Magellan/FIRE & 2.20 & 1.27 & 14.6 & 18.4 & L112-1333 & --- \\
           &  &  & g'r'i'z' & SDSS & --- & --- & --- & --- & --- & Juri{\'{c}} et al. \\
60546 & 2013 02 10 & 21 & IR & Magellan/FIRE & 2.42 & 1.45 & 5.3 & 17.8 & 2MASS J11275215-1045394 & --- \\
           & 2013 02 15 &  & g'r’i'z’ & KPNO2.1 & 2.43 & 1.45 & 5.1 & 17.8 & SDSS field & --- \\
63440 & 2012 10 19 & 32 & IR & IRTF/SpeX & 1.81 & 0.84 & 9.6 & 16.7 & L93-101 & --- \\
           & 2012 11 09 & 94 & IR & IRTF/SpeX & 1.80 & 0.83 & 10.9 & 16.8 & L93-101 & --- \\
           & 2012 11 10 & 32 & IR & IRTF/SpeX & 1.80 & 0.84 & 11.5 & 16.8 & L93-101 & --- \\
           & --- &  & g'r'i'z' & SDSS & --- & --- & --- & --- & --- & Juri{\'{c}} et al. \\
74096 & 2013 10 13 & 64 & IR & IRTF/SpeX & 2.26 & 1.27 & 5.1 & 18.2 & L115-271 & ---\\
           &            &    &    &           &      &      &      &      & L93-101 & \\
88604 & 2013 06 12 & 40 & IR & IRTF/SpeX & 2.88 & 1.88 & 4.0 & 17.3 & L105-56 & --- \\
           &            &    &    &           &      &      &      &      & L107-998 & \\
92652 & 2013 03 06 & 28 & IR & IRTF/SpeX & 2.31 & 1.32 & 2.8 & 18.2 & L98-978 & --- \\
           & 2010 03 08 &  & BVRI & Magellan/IMACS & 2.47 & 2.00 & 22.8 & 20.1 & --- & Moskovitz \\
101703 & 2013 10 03 & 40 & IR & IRTF/SpeX & 2.38 & 1.40 & 6.8 & 18.2 & L112-1333 & --- \\
             &            &    &    &           &      &      &      &      & L93-101 &
\enddata
\label{tab:ObsCircum}
\end{deluxetable*}

\begin{deluxetable}{llll}
\tablecolumns{4}
\tablewidth{0pt}
\tablecaption{Solar magnitude for normalization of asteroids' visible colors}
\tablehead{
\colhead{Filter} &
\colhead{Mag} &
\colhead{Filter} &
\colhead{Mag}
}
\startdata
{\it g'} & 5.12 & {\it B} & 5.47 \\
{\it r'} & 4.68 & {\it V} & 4.82 \\
{\it i'} & 4.57 & {\it R} & 4.46 \\
{\it z'} & 4.54 & {\it I} & 4.14
\enddata
\label{tab:SolarMagCalib}
\end{deluxetable}

\section{Analysis}
\label{sec:analysis}

\subsection{Age Calculation}
\label{sec:ageCalculation}

	The time passed since the progenitor's fission into the asteroid pair is referred to here as the pair's age. We used methods previously introduced by Vokrouhlick{\'y} \& Nesvorn{\'y} (2008, 2009) to estimate ages of the selected pairs in this study and outline the steps here (see also Supplementary materials in Pravec et al. 2010).

	We perform backward orbital integrations of multiple clones for each of the pair’s components and search their close approaches in the past. The clones are twofold in nature: (i) a first class describes the orbital uncertainty as it follows from the orbit determination based on the available astrometric data, and (ii) the second class takes into account different strength of the thermal (Yarkovsky) forces on the asteroids. The orbits are propagated backward in time to maximum of $\sim2 My$ (beyond which the orbit integrations are deemed too uncertain). The estimated asteroid sizes provide us with a quantitative basis for the convergence distance of the pair components: the asteroids are required to approach at least to a Hill sphere distance of the progenitor object and at a speed which is much smaller than its escape velocity. These criteria are always used in the results reported below.

\begin{figure}
\centerline{\includegraphics[width=8.5cm]{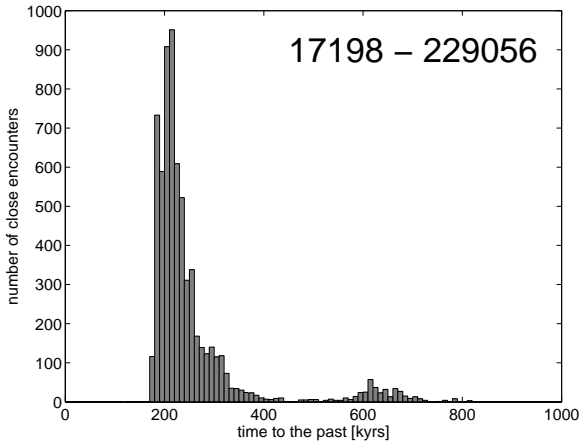}}
\caption{The ages of the pairs are estimated by backward integration of clones of the pairs' components. This is an example for the pair consisting of the primary asteroid 17198 and a secondary component 229056. We consider close encounters between couples of the clones and characterize their distribution in incremental (this figure) and cumulative way.
\label{fig:Fig2}}
\end{figure}

	The nature of our method implies that we cannot pinpoint a unique age solution for the pairs. Rather, we are left with a statistical evaluation of a multitude of possible solutions given by a cross-check between the state vectors of different clones. We typically set $\sim5-10$ ky bins in time and determine how many cases converged in a given time bin in the past. The accuracy of the solution critically depends on several factors: (i) precision of the orbit determination for the asteroids in the pair (determined by the number of observations and arc length they cover), (ii) size of the asteroids in the pair (especially since smaller secondaries are subject to a strong and unconstrained Yarkovsky effect), (iii) location of the pair in the main belt (since more chaotic regions triggered by resonances result in rapid loss of orbital predictability). A favorably accurate solution may lead to an incremental age distribution symmetric about some central value T (see Fig.~\ref{fig:Fig2} for an example) with a standard deviation $\Delta T$ (in these cases we can simply report the age solution by $T \pm \Delta T$). More often, though, the incremental age distribution is not symmetric but skewed toward older ages. In these cases we compute a cumulative distribution of converging ages and evaluate the median time and the times when $5\%$ and $95\%$ cases converged. This would give us an asymmetric interval of ages around the median value. A similar way of reporting the age was also used in Pravec et al. (2010). In most cases, our results are consistent with previous solutions. However, some are slightly different, because new astrometry makes the orbits more accurate and eventually helps the solutions shrink their uncertainty.

\subsection{Principal Component Analysis}
\label{sec:PCA}

	The derived spectra were compared to the Bus-DeMeo taxonomy to find the best match using the Principal Component Analysis (PCA) method (DeMeo et al. 2009). We used a batch-mode program that is equivalent to the web-service tool available on http://smass.mit.edu/busdemeoclass.html. PCA reduces the multidimensional input (the reflectance values at each wavelength) by transforming the data in a way that maximizes the variance along a single axis, referred to as Principal Component 1 (PC1'). After removing the variance of PC1', the method retransforms the data along a new axis, named PC2', and so on. The first few principal components contain most of the variance. Prior to the PCA calculation, the program creates a ``spline fit" to smooth the reflectance and removes its slope. This is done to avoid contamination of the resulting data by noise, missing data, bad atmospheric calibration or weathering effects. As an output, the program presents the reflectance slope, PC1' to PC5' values, the best taxonomic matches to the data and chi2 values of each matching. Since we do not have the visible data for all the pairs, we performed the PCA on the IR range for all of the 31 objects. In order to distinguish between S-, Sq and Q-type reflectance spectra we also run the PCA on the IR and visible range for the S-complex asteroids. For the six cases without observed data in the visible regime we run the PCA twice: with a visible reflectance of the S-type archetype, and the visible reflectance of the Q-type archetype, both from DeMeo et al. (2009).

\subsection{Spectral Slope and Band Analysis}
\label{sec:bandAnalysis}

	While matching the reflectance spectrum to the Bus-DeMeo taxonomy allows us to distinguish between the S-complex and other groups (e.g., C-, X-, V-, A-types), there are a few questions this method cannot address. The main problem is that the extent of weathering does not reveal itself just by the letters of the S-, Sq- and Q-types taxonomies. While in recent years it is accepted that S-type stands for the most weathered spectrum, Q-type for the most fresh, and Sq-type is an intermediate phase (Binzel et al. 1996, Binzel et al. 2004, Chapman 2004, DeMeo et al. 2009, Binzel et al. 2010, Dunn et al. 2013, and many others), it should be stressed that this is not always the case. These divisions were formed by averaging many spectra that were not necessarily identical, thus every type has some variation. In addition, the borders between the taxonomic groups are arbitrary (DeMeo et al. 2009). Moreover, the division between S-, Sq- and Q-types also includes some mineralogical differences (Q-types are richer with olivine; Gaffey et al. 1993) due to the fact that Q-types were defined by NEAs that have higher ratio of olivine-rich OC (Vernazza et al. 2008).

\begin{figure}
\centerline{\includegraphics[width=8.5cm]{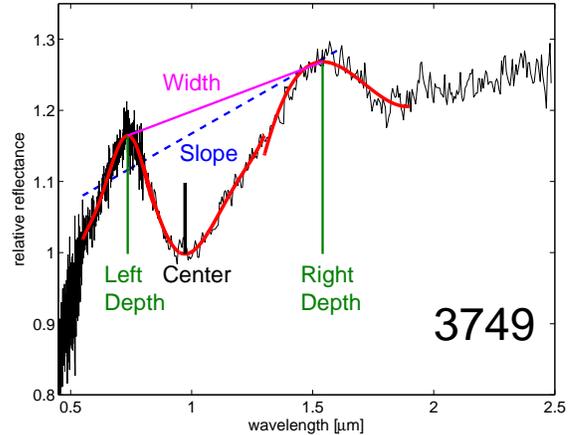}}
\caption{The band analysis used in our study: the red lines are the fits to the minima and peaks of the $1 \mu m$ absorption band. This example is for the asteroid 3749.
\label{fig:Fig3}}
\end{figure}

	Since the main effect of the space weathering mechanism on asteroidal spectra is the increasing of spectral slope and reducing of band depth (Clark et al. 2002, Brunetto et al. 2006), we analyzed the reflectance spectra focusing on the spectral slope and on the absorption band at 1 $\mu m$ of all S-complex spectra. The literature consists of many different ways for slope calculation and band analysis (e.g., Gaffey et al. 1993, Vernazza et al. 2008, Thomas \& Binzel 2010, DeMeo et al. 2014), but we tried to use an analysis method that will be the most effective for our data that is mostly infrared. In addition, we did not analyze the parameters of the 2 $\mu m$ absorption band since in many cases it was too noisy to derive significant results (therefore we did not calculate the ``Band Area Ratio", BAR; Gaffey et al. 1993). The formalism of this band analysis includes (See Fig.~\ref{fig:Fig3} for an illustrated explanation):

{\it i)	} A linear fit to the spectrum from 0.55 to 1.6 $\mu m$ is defined as the Spectral Slope. These values were used since all spectra are normalized to unity at 0.55 $\mu m$, therefore this value could be used as a common base for different spectra while the difference in slope between the weathered S-type to the fresh Q-type asteroids is maximized at $\sim1.6 \mu m$. In addition, after 1.6 $\mu m$ the spectra are usually noisier (telluric lines start to appear around 1.8 $\mu m$).

{\it ii)} A three to five order polynomial fit to the band minima between 0.8 to 1.3 $\mu m$. The wavelength at the minima is the Band Center. This allowed us to measure a secure center even for those asteroids without measured visible observations.

{\it iii)} A three to five order polynomial fit to the maxima between 0.55 to 0.85 $\mu m$. The wavelength at the maxima of the fit is referred as the Left Peak Wavelength.

{\it iv)} A three to five order polynomial fit to the maxima between 1.4 to 1.8 $\mu m$. The wavelength at the maxima of the fit is referred as the Right Peak Wavelength.

{\it v)} The Band Width is defined as the distance between the reflectance values at the Left Peak to the one at the Right Peak Wavelength.

{\it vi)} The orthogonal distance (parallel to the y-axis) between the reflectance value of the Left Peak Wavelength and the reflectance value at the Band Center, is defined as the band's Left Depth.

{\it vii)} The orthogonal distance (parallel to the y-axis) between the reflectance value at the Right Peak Wavelength and the reflectance value at the Band Center, is defined as the band's Right Depth.

\begin{figure}
\centerline{\includegraphics[width=8.5cm]{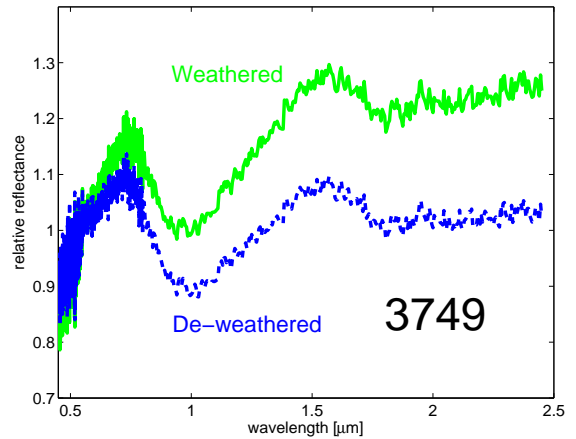}}
\caption{De-weathering example for the spectrum of 3749. The original spectrum (green line) is modified by Eq. (1) in a recursive way until the slope is $0.03\%$ per $\mu m$ (blue dashed line). This value is the average slope of the spectra of OC meteorites.
\label{fig:Fig4}}
\end{figure}

{\it viii)} We applied the formula (Eq. 1) of Brunetto et al. (2006) to ``de-weather" the reflectance spectra (Fig.~\ref{fig:Fig4}). Then we applied the band analysis (sections {\it ii} - {\it vii}) in order to derive the ``original" band parameters of the asteroids.

	The uncertainty of these parameters is based on {\it 1)} the SNR of the spectrum (random error) and {\it 2)} on the systematic errors that result from the observational conditions (airmass, weather), and the solar analog that is being used for the calibration. Therefore, to estimate the random error we added to the reflectance value a randomized noise that is in the order of the measured reflectance noise. We saved the new reflectance spectrum and delivered it as an input to our analysis code. We repeated this algorithm a thousand times and stored the standard deviation of the thousand artificial spectra as the uncertainty of each of the spectral parameters. We estimated the systematic error by measuring the band parameters of different observations of the same asteroid (we used multiple observations of 2110 and 3749) and calculated their standard deviation. We used the larger value between the random and the systematic errors as the uncertainty of the band parameters. When the measured visible spectrum was not accessible, we separately analyzed the spectral parameters of the two extreme cases of S-type visible spectra and Q-type visible spectra as described above, then averaging the two results to get a single value per asteroid, but using the two results as the limit of the uncertainty range.

\section{Results}
\label{sec:results}

\subsection{Objects}
\label{sec:objects}

\begin{figure}
\centerline{\includegraphics[width=8.5cm]{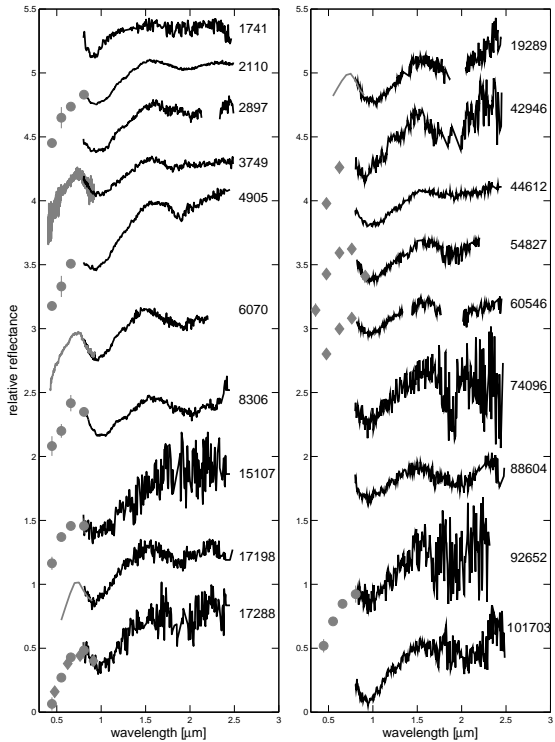}}
\caption{Reflectance spectra of asteroid pairs of the S-complex taxonomy. The spectra were shifted on the Y-axis for clarity with no weathering correction yet applied. Visible regime is marked in gray: {\it BVRI} colors (circles), SDSS {\it g'r'i'z'} colors (diamonds), spectra (line).
\label{fig:Fig5}}
\end{figure}

\begin{figure}
\centerline{\includegraphics[width=8.5cm]{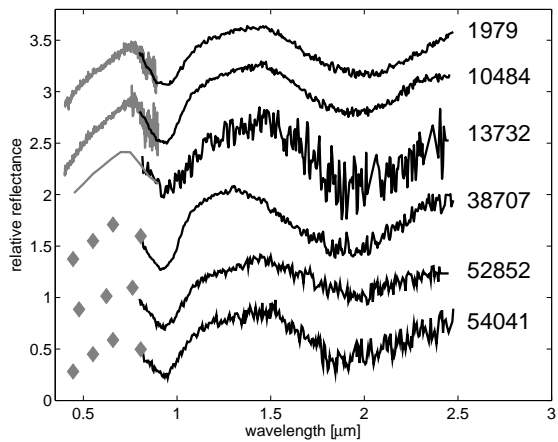}}
\caption{Reflectance spectra of asteroid pairs of the V-type taxonomy. The spectra were shifted on the Y-axis for clarity. The visible colors presented here for 38707 and 52852 belong to their secondaries 32957 and 250322, respectively. All six asteroids have an average right depth of $0.66\pm0.08$ in reflectance units between their $0.9 \mu m$ minima to their $1.4\mu m$ maxima, which is low compared to the average right depth of all V-types which is $0.8\pm0.1$.
\label{fig:Fig6}}
\end{figure}

\begin{figure}
\centerline{\includegraphics[width=8.5cm]{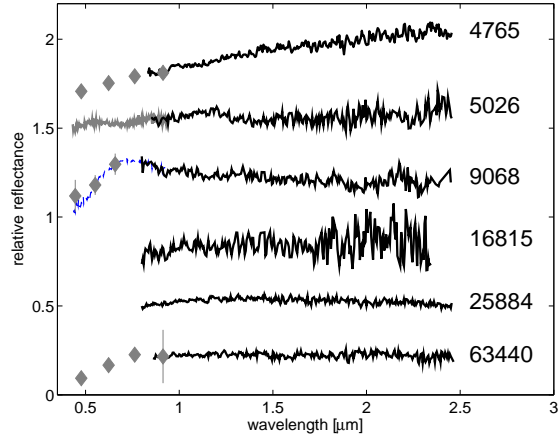}}
\caption{Reflectance spectra of asteroid pairs of the C/X-complex taxonomy. The spectra were shifted on the Y-axis for clarity. 434 Hungaria's visible spectrum (dashed line) was used to scale the BVR measurements of 9068 to its IR spectrum. 4765, 9068, 25884 and 63440 are all part of the Hungaria-family ($1.8<a<2$ AU) and they are probably Xe-type (Bus-DeMeo taxonomy) or E-type (Tholen taxonomy). A feature in the spectrum of 5026 at 0.5 to 0.7 $\mu m$ suggests that it is a Ch-type in the Bus-DeMeo taxonomy.
\label{fig:Fig7}}
\end{figure}

	The spectral reflectance of 31 asteroids in pairs were collected (Fig.~\ref{fig:Fig5}$-$~\ref{fig:Fig7}). Excluding telescope availability, we only limit target selection by a visible limiting magnitude of 18.5 for the IRTF and 19.5 for Magellan. This limiting magnitude allowed us to reach a median SNR of approximately 25 that was enough to measure the parameters of the 1 $\mu m$ absorption band of the S-complex spectra.

	20 of the observed asteroids are located in the inner main-belt ($2.0<a<2.5$ AU), six in the middle main-belt ($2.5<a<2.8$ AU) and one in the outer main-belt ($a>2.8$ AU). This is clearly an observational bias since the small drifting secondaries of asteroid pairs are hard to discover at larger distances, and without the knowledge of their existence an asteroid pair cannot be identified even if the primary member is known. In addition, four members of the Hungaria family ($1.8<a<2$ AU) were also observed.

	The diameters of the observed asteroids range from 1.8 to 14.9 km, that corresponds to an absolute magnitude range of $11.4<H<15.5$ mag. This size range fits to the expected formation model of asteroid pairs (Pravec et al. 2010), that involve objects small enough to be significantly altered by the YORP effect in a timescale of approximately $10^7$ years (Kaasalainen et al. 2007, {\v{D}}urech et al. 2008, 2012) and large enough to have a shattered, aggregate-based, ``rubble-pile" structure (Pravec \& Harris 2000), that could disintegrate due to a fast spin. Two of the observed asteroids (1741 and 16815) might be too large for an asteroid pair (primary diameter of $\sim15$ km) that are formed by the size-dependent YORP effect. However, our calculations show that the orbit of their ``partners" converge with their orbits in the last million of years and therefore we cannot rule out that these are not asteroid pairs.

	The size ratios of the secondaries to the primaries (D2/D1) of the observed asteroids range from a value of 0.76 (almost identical sizes) to 0.1 (a small secondary). Assuming a mass ratio $q = (D2/D1)^3$, this range can be translated to a range of mass ratios of 0.44 to 0.002 that spans the entire range of known asteroid pairs. The mass ratio of one pair (1979-13732) is significantly higher ($\sim 0.4$) compared to the mass ratio of the other asteroids (averaged at 0.1 and lower than 0.25). Models (e.g. Scheeres 2007) predict that there is not enough free energy in the system of disrupting asteroid to allow such a massive secondary to escape (the limit was shown to be around 0.2; Scheeres 2007, Pravec et al. 2010). Therefore, 1979's secondary (13732) might have been formed by another mechanism (such as a separation of an unstable binary asteroid). Alternatively, the absolute magnitude H of this pair might be wrong. We should note that the similarity between the spectra of these two asteroids (both V-types) supports a common origin. An asteroid pair or not, since this asteroid is not part of the S-complex group, it is omitted from our analysis anyway.

	We obtained a wide range of ages for the observed asteroids running from $17.0\pm0.5$ kyr to ages with minimum limit of 2,000 kyr and a maximal limit beyond the maximum orbital integration time we used (2 Myr).

	Physical details of the observed asteroids are detailed in Table~\ref{tab:PairsParam}.

\begin{deluxetable*}{ccccccccccccc}
\tablecolumns{13}
\tablewidth{0pt}
\tablecaption{Pairs' physical properties and age. The data for four secondary pairs appear just after their relevant primaries}
\tablehead{
\colhead{Name} &
\colhead{1/2} &
\colhead{a} &
\colhead{$H_v$} &
\colhead{D} &
\colhead{Spin} &
\colhead{Amp} &
\colhead{Partner} &
\colhead{${\Delta H}$} &
\colhead{Age} &
\colhead{Taxonomy} &
\colhead{PCir1'}          &
\colhead{PCir2'}          \\
\colhead{}          &
\colhead{}          &
\colhead{[AU]} &
\colhead{}          &
\colhead{[km]} &
\colhead{[hours]}          &
\colhead{[mag]}    &
\colhead{}          &
\colhead{[mag]}    &
\colhead{[kyrs]} &
\colhead{} &
\colhead{} &
\colhead{} \\
}
\startdata
 \hline
1741   & 1 & 2.89 & 11.4 & 14.9  & 2.94  & 0.1  & 258640    & 4.3 & 160-40+150   & S/Sq & 0.071  & 0.054  \\ \hline
1979   & 1 & 2.37 & 13.6 & 4.2   & 7.52  & 0.21 & 13732     & 0.7 & $>2000$      & V    & 0.782  & -0.059 \\
13732  & 2 & 2.37 & 14.3 & 3.1   & 8.30  & 0.28 & 1979      & 0.7 & $>2000$      & V    & 0.878  & -0.196 \\ \hline
2110   & 1 & 2.20 & 13.2 & 6.5   & 3.34  & 0.45 & 44612     & 2.3 & $>1600$      & S    & -0.061 & -0.083 \\
44612  & 2 & 2.20 & 15.5 & 2.3   & 4.91  & 0.44 & 2110      & 2.3 & $>1600$      & Sq/Q & -0.169 & -0.056 \\ \hline
2897   & 1 & 2.25 & 13.2 & 6.5   & 2.60  & 0.15 & 182259    & 3.7 & 340-150+350  & S/Sq & -0.070 & -0.141 \\ \hline
3749   & 1 & 2.24 & 13.1 & 6.8   & 2.80  & 0.14 & 312497    & 4.4 & 280-25+45    & Sq   & -0.166 & -0.084 \\ \hline
4765   & 1 & 1.95 & 13.7 & 3.7   & 3.63  & 0.56 & 350716    & 3.8 & 170-30+430   & X/E  & -0.166 & 0.142  \\ \hline
4905   & 1 & 2.60 & 12.1 & 10.8  & 6.05  & 0.41 & 7813      & 1   & $>1650$      & Sw   & -0.036 & -0.129 \\ \hline
5026   & 1 & 2.38 & 13.8 & 9.6   & 4.42  & 0.49 & 2005WW113 & 4   & $18\pm1$     & Ch   & -0.026 & 0.280  \\ \hline
6070   & 1 & 2.39 & 13.7 & 5.2   & 4.27  & 0.41 & 54827     & 1.6 & $17\pm0.5$   & Sq   & 0.023  & -0.149 \\
54827  & 2 & 2.39 & 15.3 & 2.5   & 5.88  & 0.25 & 6070      & 1.6 & $17\pm0.5$   & Q    & -0.139 & -0.094 \\ \hline
8306   & 1 & 2.24 & 14.9 & 3.0   & 3.60  & 0.1  & 2011SR158 & 3.2 & 400-100+250  & Sq   & -0.121 & -0.102 \\ \hline
9068   & 1 & 1.82 & 13.5 & 4.0   & 3.41  & 0.20 & 2002OP28  & 4.3 & 32-1+15      & X/E  & -0.214 & 0.199  \\ \hline
10484  & 1 & 2.32 & 13.8 & 3.8   & 5.51  & 0.21 & 44645     & 1   & 310-80+210   & V    & 0.827  & -0.033 \\ \hline
15107  & 1 & 2.27 & 14.3 & 3.9   & 2.53  & 0.14 & 291188    & 2.6 & 650-220+1000 & S?   & -0.239 & -0.063 \\ \hline
16815  & 1 & 2.56 & 12.6 & 10-18 & 2.9   & 0.20 & 2011GD83  & 4.7 & 95-20+40     & C/X  & -0.232 & 0.143  \\ \hline
17198  & 1 & 2.28 & 14.9 & 3.0   & 3.24  & 0.13 & 229056    & 2.6 & 230-50+120   & Sw   & 0.029  & -0.088 \\ \hline
17288  & 1 & 2.29 & 14.1 & 4.3   & 4.3   & 0.15 & 203489    & 2.3 & 700-180+520  & Sw   & -0.163 & -0.110 \\ \hline
19289  & 1 & 2.12 & 15.3 & 2.5   & 2.85  & 0.16 & 278067    & 2.3 & 1250-100+400 & Q    & -0.156 & -0.054 \\ \hline
25884  & 1 & 1.95 & 14.6 & 2.4   & 4.92  & 0.55 & 48527     & 1.5 & 420-100+200  & X/E  & -0.109 & 0.159  \\ \hline
38707  & 1 & 2.28 & 14.9 & 2.3   & 6.15  & 0.36 & 32957     & 1.1 & $>2000$      & V    & 1.287  & 0.249  \\ \hline
42946  & 1 & 2.57 & 13.6 & 5.4   & 3.42  & 0.30 & 165548    & 2.1 & 600-150+580  & Srw  & 0.193  & 0.024  \\ \hline
52852  & 1 & 2.26 & 14.8 & 2.4   & 5.43  & 0.19 & 250322    & 2   & 330-30+800   & V    & 0.685  & -0.119 \\ \hline
54041  & 1 & 2.32 & 14.5 & 2.8   & 18.86 & 0.23 & 220143    & 2   & 150-30+470   & V    & 0.864  & -0.035 \\ \hline
63440  & 1 & 1.94 & 15.2 & 1.8   & 3.30  & 0.17 & 331933    & 2.2 & 33-4+17      & X/E  & -0.152 & 0.192  \\ \hline
74096  & 1 & 2.38 & 15.5 & 2.3   & 5.99  & 0.27 & 224857    & 1.5 & 320-150+750  & S/Sq & -0.131 & -0.237 \\ \hline
88604  & 1 & 2.67 & 13.3 & 6.2   & 7.18  & 0.55 & 60546     & 1.3 & $>1000$      & S/Sq & 0.004  & -0.017 \\
60546  & 2 & 2.67 & 14.6 & 3.4   & ---   & ---  & 88604     & 1.3 & $>1000$      & S    & -0.063 & -0.023 \\ \hline
92652  & 1 & 2.34 & 15.5 & 2.3   & ---   & ---  & 194083    & 1.3 & 100-30+1050  & Sw   & -0.062 & -0.072 \\ \hline
101703 & 1 & 2.54 & 15.1 & 2.7   & 3.90  & 0.29 & 142694    & 2.0 & 600-150+100  & Sw/Q & 0.088  & -0.113
\enddata
\tablenotetext{}{Source of data:}
\tablenotetext{-}{Age, taxonomy, PCir1' and PCir2' were measured and calculated in our study.}
\tablenotetext{-}{Semi-major axis, absolute magnitudes, and $\Delta H$ are from the MPC website.}
\tablenotetext{-}{Diameters were estimated from the absolute magnitude assuming an albedo value of 0.22 for S-complex asteroids, 0.36 for V-type, 0.43 for E-type, 0.05 to 0.15 for C/X-complex and 0.058 for the Ch-type asteroid (Mainzer et al. 2011).}
\tablenotetext{-}{Rotation periods and amplitude are taken from Polishook 2011 (25884), Polishook et al. 2011 (3749), Polishook 2014 (4905, 8306, 17288, 16815, 42946, 74096), Pravec et al. 2010 (2110, 4765, 5026, 6070, 10484, 13732, 15107, 17198, 19289, 38707, 44612, 52852, 54041, 54827, 63440, 88604, 101703), Slivan et al. 2008 (1741), Warner et al. 2009 (9068) and Pravec's web-page: http://www.asu.cas.cz/$\sim$ppravec (1979, 2897).}
\tablenotetext{-}{``Partner" (usually secondary) information is from Pravec and Vokrouhlick{\'y} (2009), Vokrouhlick{\'y} (2009), Ro{\.{z}}ek et al. (2011) and Pravec (private communication, 2013) who identified the asteroid pairs involving the primaries 8306, 4905, 16815 and 74096 with the method of Pravec and Vokrouhlick{\'y} (2009), updated with the use of mean elements from AstDyS-2.}
\label{tab:PairsParam}
\end{deluxetable*}

\subsection{Taxonomy and Principal Component Analysis}
\label{sec:taxonomy}

\begin{figure}
\centerline{\includegraphics[width=8.5cm]{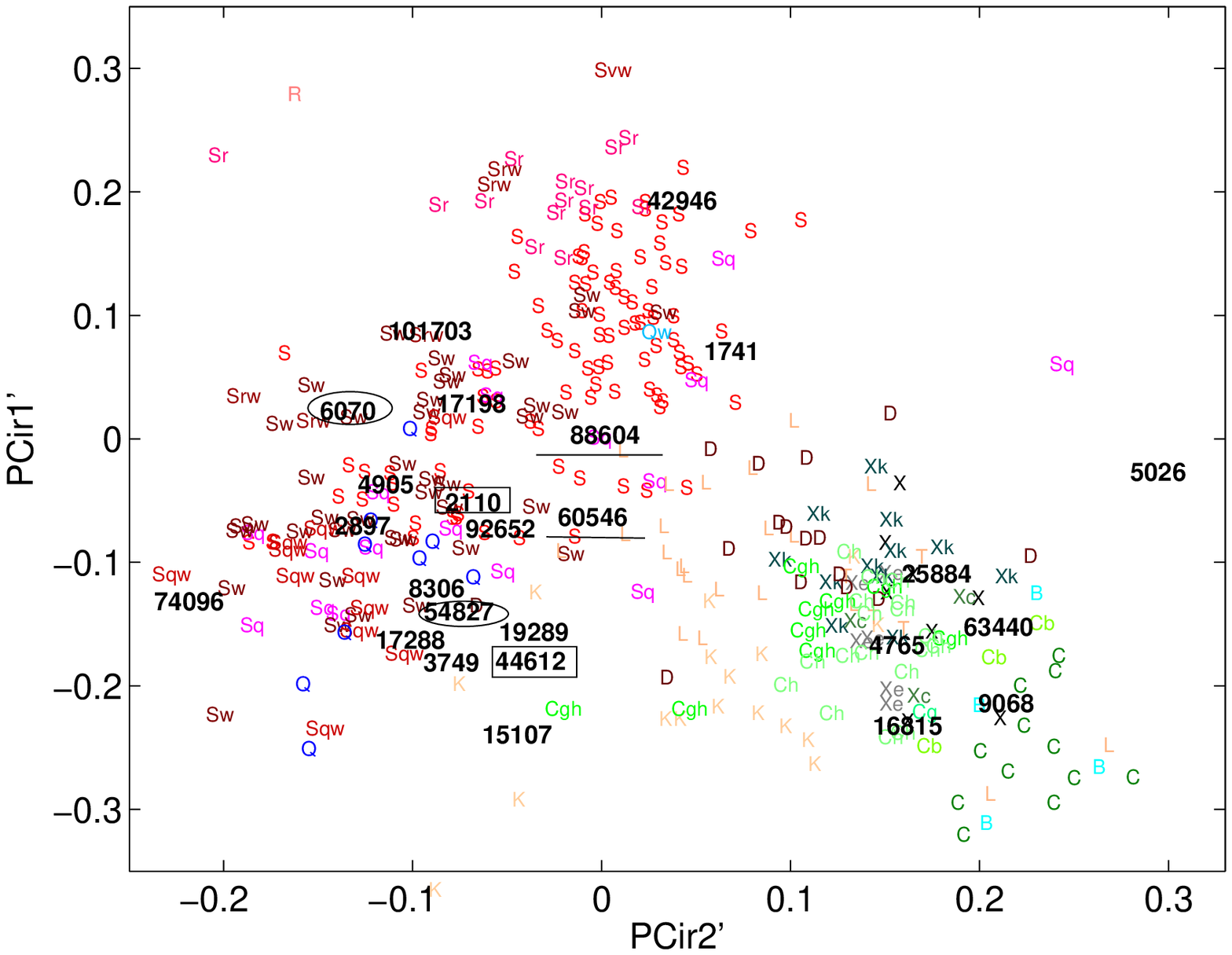}}
\caption{PCir2' vs PCir1' values from the IR spectra of the asteroid pairs (marked numbers) and the background population (taxonomic letters) taken from DeMeo et al. (2009). The primary and secondary which belong to the same pair are marked (2110-44612 - underline; 6070-54827 - ellipse; 88604-60546 - rectangle). Some pairs were slightly shifted in order not to overwrite their numbers (exact values appear in Table~\ref{tab:PairsParam}). The displayed section includes the S- and C/X-complexes.
\label{fig:Fig8a}}
\end{figure}

\begin{figure}
\centerline{\includegraphics[width=8.5cm]{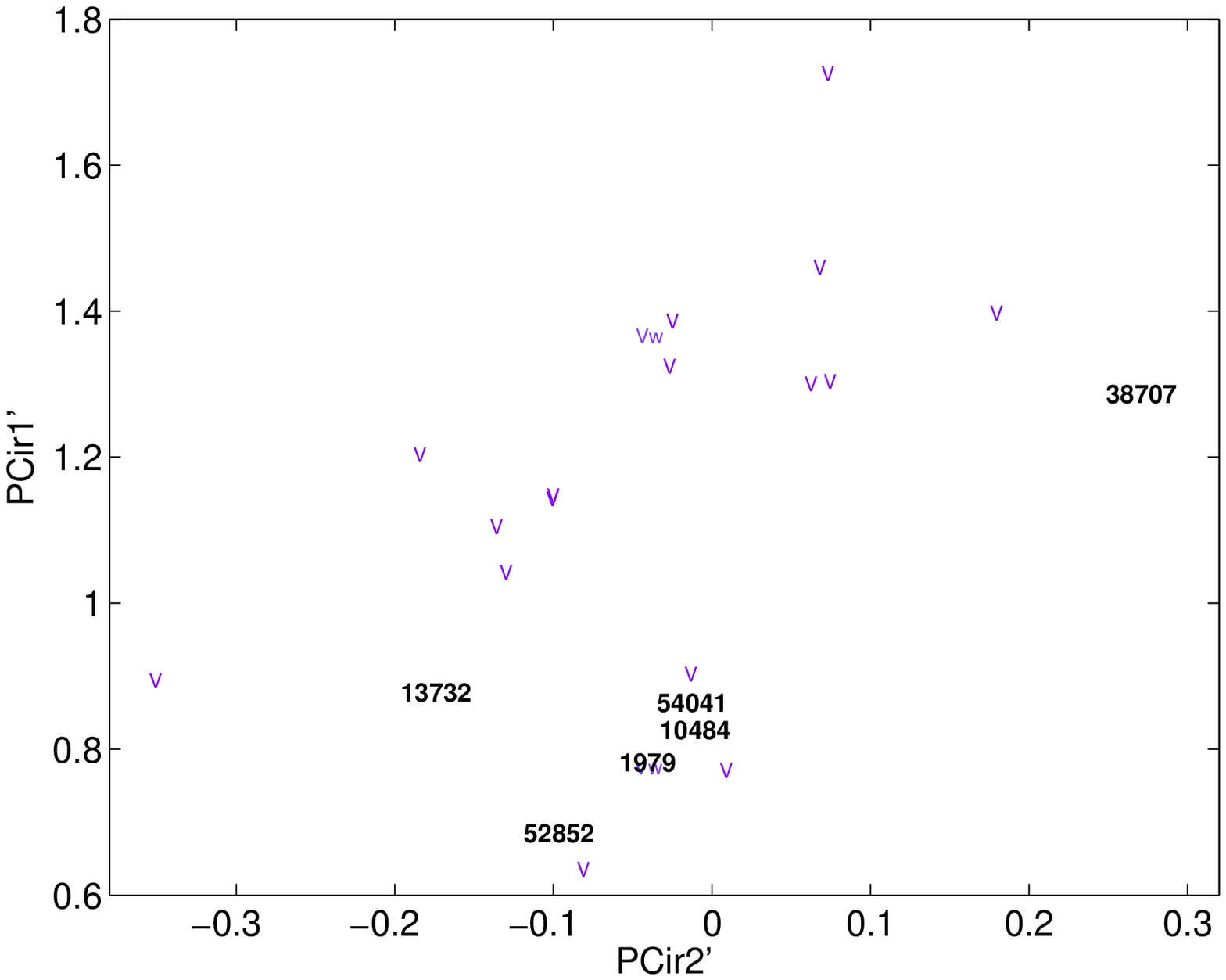}}
\caption{PCir2' vs PCir1' values from the IR spectra of the asteroid pairs (marked numbers) and the background population (taxonomical letters) taken from DeMeo et al. (2009). The displayed section includes V-type asteroids. Values appear in Table~\ref{tab:PairsParam}. V-type pairs tend to clump in the lower section of this plot, which indicates low values of the band's right depth. We stress this result even though we cannot explain why V-type pairs tend to have low-minima right depth, or if this behavior is actually representative of all V-type pairs.
\label{fig:Fig8b}}
\end{figure}

	Figures~\ref{fig:Fig8a} and ~\ref{fig:Fig8b} present the main two vectors of the principal components analysis for the infrared range, PCir2' vs. PCir1' of the asteroid pairs and the background population from DeMeo et al. (2009) that defines the taxonomy. The PCir values of the pairs are presented in Table~\ref{tab:PairsParam}. 19 of the pairs are located in the S-complex area, 6 in the C/X area and 6 in V-type area.

\begin{figure}
\centerline{\includegraphics[width=8.5cm]{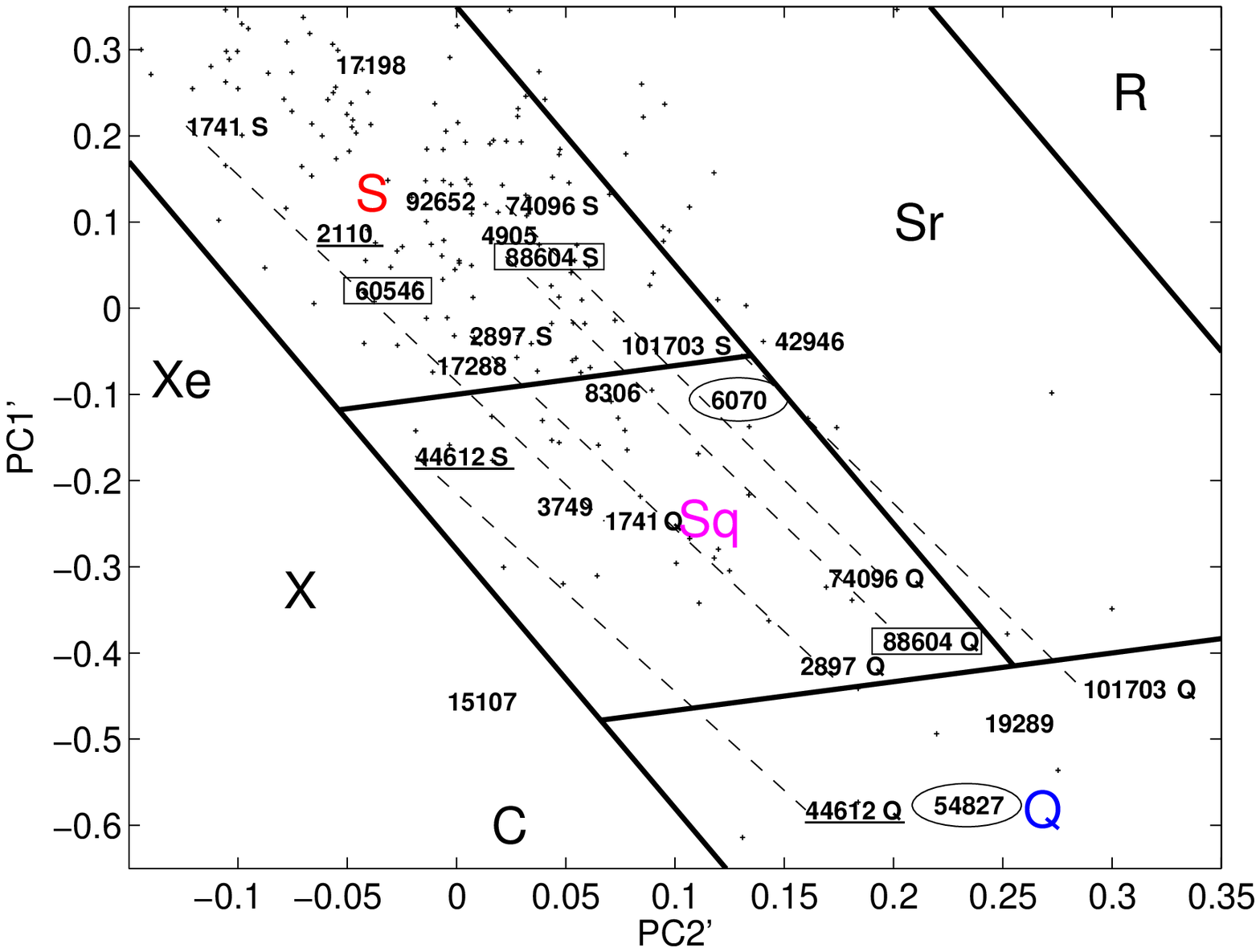}}
\caption{PC2' vs PC1' values from the visible+IR spectra of the observed asteroid pairs of the S-complex taxonomy (marked numbers) and the background population of S-complex (pluses; taken from DeMeo et al. 2009). These values show that 2 of the pairs are Q-types, 2 are Sq/Q, 3 are Sq, 4 are Sq/S and 7 are S-types. The division between the taxonomies (S-, Sq-, Q-, Sr-, R-type) is from DeMeo et al. (2009). Six asteroids that were not observed in the visible appear twice: once with a S-type visible spectrum (marked by a suffix ``S"), and one with a Q-type visible spectrum (marked by a suffix ``Q"); Dashed-lines connect their instances. The primary and secondary which belong to the same pair are marked (2110-44612 - underline; 6070-54827 - ellipse; 88604-60546 - rectangle). Some of the pairs were slightly shifted in order not to overwrite their numbers (the exact values appear in Table~\ref{tab:BandAnalysis}). While 15107 is off the defined S-complex area in the PC2'-PC1' plane it is defined as S-complex by the PCir2'-PCir1' plain, and it seems to have an absorption band around 1 $\mu m$ as S-complex asteroids have. Therefore, we consider it as part of the group.
\label{fig:Fig9}}
\end{figure}

	Since the PCA of the IR reflectance alone cannot distinguish between S-, Sq- and Q-types we further run the PCA on the full infrared and visible range of the 19 S-complex pairs. The results (Fig.~\ref{fig:Fig9}) show that 2 of the pairs are Q-types, 2 are Sq/Q, 3 are Sq, 4 are Sq/S and 7 are S-types. One object, 15107, falls outside of the S-complex area on the Visible and infrared PCA although it falls in the S-complex area on the infrared PCA alone, probably due to low S/N of the observations. Since its 1 $\mu m$ absorption band is noticeable, we consider it as an S-complex asteroid.

\begin{deluxetable*}{cccccccc}
\tablecolumns{8}
\tablewidth{0pt}
\tablecaption{Spectral analysis of the S-complex pairs}
\tablehead{
\colhead{Name} &
\colhead{PCvisir1'} &
\colhead{PCvisir2'} &
\colhead{Slope} &
\colhead{Band 1} &
\colhead{Band 1} &
\colhead{Band 1} &
\colhead{Band 1} \\
\colhead{} &
\colhead{} &
\colhead{} &
\colhead{} &
\colhead{depth} &
\colhead{center} &
\colhead{de-weathered center} &
\colhead{width} \\
}
\startdata
1741   & 0.211 - -0.247  & -0.124 - 0.068 & $0.20\pm0.08$ & $0.18\pm0.03$ & $0.90\pm0.01$ & $0.91\pm0.01$ & $0.79\pm0.05$ \\
2110   & 0.087           & -0.064         & $0.40\pm0.08$ & $0.07\pm0.02$ & $0.95\pm0.01$ & $0.99\pm0.01$ & $0.81\pm0.05$ \\
44612  & -0.172 - -0.582 & -0.019 - 0.160 & $0.16\pm0.08$ & $0.15\pm0.03$ & $0.99\pm0.01$ & $1.01\pm0.01$ & $0.85\pm0.11$ \\
2897   & -0.032 - -0.440 & 0.006 - 0.177  & $0.32\pm0.08$ & $0.13\pm0.04$ & $0.96\pm0.01$ & $0.99\pm0.01$ & $0.91\pm0.05$ \\
3749   & -0.23           & 0.037          & $0.19\pm0.08$ & $0.17\pm0.02$ & $0.97\pm0.01$ & $0.99\pm0.01$ & $0.81\pm0.05$ \\
4905   & 0.065           & 0.011          & $0.59\pm0.08$ & $0.13\pm0.02$ & $0.91\pm0.01$ & $0.93\pm0.01$ & $0.93\pm0.05$ \\
6070   & -0.106          & 0.122          & $0.26\pm0.08$ & $0.20\pm0.02$ & $0.96\pm0.01$ & $0.98\pm0.01$ & $0.79\pm0.05$ \\
54827  & -0.577          & 0.219          & $0.04\pm0.08$ & $0.24\pm0.02$ & $0.98\pm0.01$ & $0.98\pm0.01$ & $0.86\pm0.05$ \\
8306   & -0.088          & 0.059          & $0.10\pm0.08$ & $0.23\pm0.02$ & $1.00\pm0.01$ & $1.00\pm0.01$ & $0.86\pm0.05$ \\
15107  & -0.456          & -0.004         & $0.35\pm0.08$ & $0.08\pm0.02$ & $0.94\pm0.01$ & $0.99\pm0.02$ & $1.12\pm0.05$ \\
17198  & 0.282           & -0.055         & $0.45\pm0.08$ & $0.14\pm0.02$ & $0.92\pm0.01$ & $0.96\pm0.01$ & $0.86\pm0.05$ \\
17288  & -0.047          & -0.009         & $0.36\pm0.08$ & $0.14\pm0.02$ & $1.00\pm0.01$ & $1.03\pm0.01$ & $0.88\pm0.05$ \\
19289  & -0.481          & 0.242          & $0.16\pm0.08$ & $0.20\pm0.02$ & $0.99\pm0.01$ & $1.01\pm0.01$ & $0.82\pm0.05$ \\
42946  & -0.038          & 0.146          & $0.47\pm0.08$ & $0.12\pm0.02$ & $0.88\pm0.01$ & $0.91\pm0.01$ & $0.83\pm0.16$ \\
74096  & 0.119 - -0.313  & 0.023 - 0.195  & $0.32\pm0.08$ & $0.15\pm0.04$ & $0.95\pm0.01$ & $0.97\pm0.01$ & $0.94\pm0.06$ \\
88604  & 0.060 - -0.386  & 0.022 - 0.205  & $0.18\pm0.08$ & $0.16\pm0.03$ & $0.93\pm0.01$ & $0.95\pm0.01$ & $0.75\pm0.05$ \\
60546  & 0.020           & -0.047         & $0.19\pm0.08$ & $0.10\pm0.02$ & $0.97\pm0.02$ & $0.99\pm0.01$ & $0.86\pm0.05$ \\
92652  & 0.124           & -0.023         & $0.53\pm0.08$ & $0.05\pm0.02$ & $0.94\pm0.02$ & $1.02\pm0.02$ & $0.81\pm0.05$ \\
101703 & -0.053 - -0.442 & 0.130 - 0.287  & $0.42\pm0.08$ & $0.17\pm0.02$ & $0.94\pm0.01$ & $0.96\pm0.01$ & $0.9\pm0.1$
\enddata
\label{tab:BandAnalysis}
\end{deluxetable*}

	Among the 6 objects that have feature-less spectra and therefore fall in the C- and X-complex area, four belong to the Hungaria group of asteroids at semi-major axis that is lower than 2.0 AU. This is not surprising since (434) Hungaria is a Tholen E-type asteroid (Tholen 1984) that partly matches the X-complex in the Bus (Bus \& Binzel 2002) and Bus-DeMeo taxonomy (DeMeo et al. 2009). Two of these pairs (4765, 25884) were measured by WISE spacecraft telescope to show albedo values larger than $20\%$ (Masiero et al. 2011) that is typical for the Enstatite-rich E-type asteroids that have an average albedo of $\sim40\%$. In addition, the B, V and R magnitude of 9068 match nicely to the visible spectrum of 434 Hungaria (Fig.~\ref{fig:Fig7}; Binzel et al. 2004). Therefore, we conclude that 4765, 9068, 25884 and 63400 are all E-type / Xe-type asteroids. The other objects with featureless spectra (5026 and 16815) are located in the inner and mid main-belt (a=2.38 and a=2.56 AU). A feature in the visible spectrum of 5026 at 0.5 to 0.7 $\mu m$ suggests that it is a Ch-type in the Bus-DeMeo taxonomy. 16815 might be a C- or X-complex and therefore the uncertainty on its diameter is large (Table~\ref{tab:PairsParam}).

	The six V-type asteroids in the pair sample are concentrated in the lower edge of the V-type area on the PCir2' - PCir1' plane. Although clearly V-types, these objects (1979, 10484, 13732, 38707, 52852, 54041) have an average right depth of $0.66\pm0.08$, which is low compared to the average value of all V-types, $0.8\pm0.1$. Vesta itself, which is unique compared to other V-types, has a right depth of 0.5 in reflectance units. We stress this result even though we cannot explain why V-type pairs tend to have low-minima right depth, or if this behavior is actually representative of all V-type pairs.

\begin{figure}
\centerline{\includegraphics[width=8.5cm]{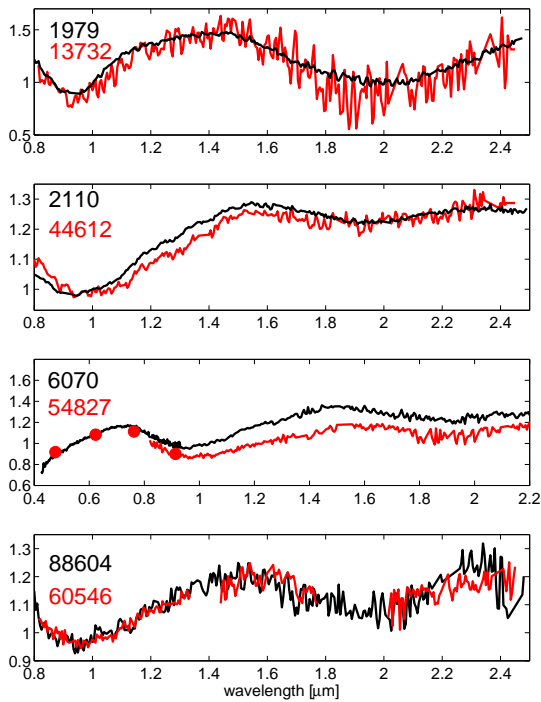}}
\caption{Reflectance spectra of primaries (black curve) and secondaries (red curve) of the same pairs. The members of each pair display a similar spectrum - this support the idea of common origin of each pair. The S-complex pairs (2110-44612, 6070-54827) present some differences in spectral slope, the marker of the weathering process - this is discussed in the text.
\label{fig:Fig10}}
\end{figure}

	We collected spectroscopic data for both members in four asteroid pairs, namely 1979-13732, 2110-44612, 6070-54827 and 88604-60546. There is a match between the spectra of the primary and the secondary in all four cases (Fig.~\ref{fig:Fig10}). This supports the idea of a shared origin of each asteroid pair. Three of the pairs belong to the same S-complex group; two of these pairs have some differences in the spectral slopes and band depths, the markers of the weathering process; these differences are discussed below.

	Moskovitz (2012) also found similarities between the primaries and secondaries by comparing the visible colors of asteroid pairs. Duddy et al. (2012, 2013) compared visible spectra of five primaries and five secondaries belonging to the same pairs and in three cases they found similarity between primaries and secondaries: 1979-13732 are both R-type, 7343-154634 are both S-type, and 11842-228747 that are both Sr-type. Mismatches were found between 17198 and 229056 and between 19289 and 278067 that were classified by Duddy et al. as A-, R-, Sr- and Q-type, respectively. However, the narrow range of visible spectra is degenerate when one tries to classify reflectance spectra by the Bus-DeMeo taxonomy that is based on visible and infrared spectroscopy. For example, the visible section (0.45 to 0.9 $\mu m$) of different types of the S-complex (S-, Sq-, Sr-, Sa-, Q-) and even A-, R- and V-types are very similar to one another especially when low signal to noise is involved. We measured the infrared spectra of four of the objects studied by Duddy et al. (2013) and found that 1979 and 13732 have a much better fit to the V-type taxonomy, 17198 has a S-type spectrum, and 19289 has a Q-type spectrum, therefore, its spectrum match to that of its secondary, 278067, that was classified by Duddy et al. (2013) as a Q-type. We re-classify the ten asteroid pairs measured by Duddy et al. as S-complex (7343-154634, 11842-228747, 17198-229056), Q-type (19289-278067) and V-type (1979-13732). An independent study by Stephen Wolters (submitted) confirms the taxonomic analysis derived in our study.

\subsection{Analysis of Spectral Slopes and Band Parameters}
\label{sec:bandParameters}

\begin{figure}
\centerline{\includegraphics[width=8.5cm]{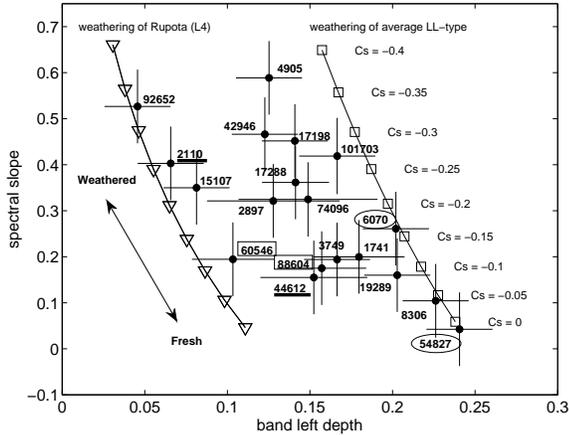}}
\caption{The pairs' spectral slope vs. their band's left depth (black circles), the main spectral parameters that are modified by space weathering. The primary and secondary belong to the same pair are marked (2110-44612 - underline; 6070-54827 - ellipse; 88604-60546 - rectangle). To show the extent of weathering we present the model of Brunetto et al. (2006) applied on an average LL-type meteorite (rectangles) and a meteorite with a low band depth (Rupota (L4) - triangles; Dunn et al. 2010). As the $C_{S}$ parameter is lower, the reflectance spectrum is more weathered, thus increasing weathering occurs up and to the left. Most importantly this analysis reveals two asteroids that appear the least weathered; 8306 and 54827.
\label{fig:Fig11}}
\end{figure}

	To reveal the weathering state of the asteroid pairs behind the taxonomic classes we plot the spectral slope vs. the left band depth (Fig.~\ref{fig:Fig11}) - the main spectral parameters that are modified by space weathering (Clark et al. 2002). To show the extent of the weathering we compared the pairs’ values to the model of Brunetto et al. (2006) (Eq. 1). As an input to the model we used a reflectance spectrum of an average LL-type meteorite that has a deep band depth and an almost zero spectral slope. Since the band depth is not only a function of the weathering but is also determined by the grain size distribution, mafic mineral abundance, opaque abundance and impact melt abundance (Reddy et al. 2012a), we also applied Eq. (1) on a meteoritic spectrum with a low band depth (Rupota (L4); Dunn et al. 2010), in order to show that the weathering trend progresses at parallel lines on the slope-depth plane.

\begin{figure}
\centerline{\includegraphics[width=8.5cm]{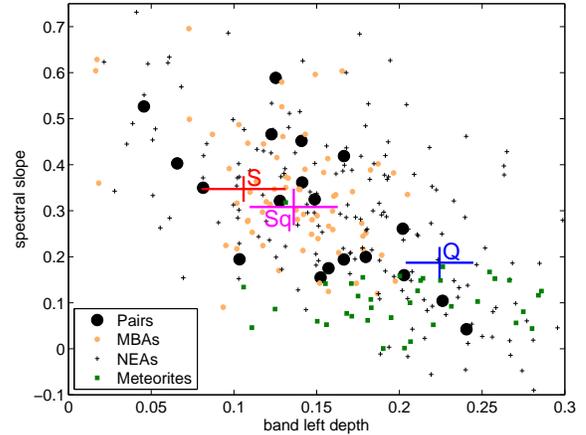}}
\caption{Same as Fig.~\ref{fig:Fig11} for the 19 observed OC pairs (black circles), 70 OC MBAs (in a size range of $1<D<15$ km, to match that of the pairs; orange dots), 178 OC NEAs (pluses) and 48 OC meteorites (green rectangles). The names and uncertainties of the pairs appear on Fig.~\ref{fig:Fig11} and were removed here for clarity. Two pairs have spectral slopes that match those of meteorites; six to seven present slope values on the edge between the weathered MBAs to the fresh meteorites. Ten to eleven have slopes values that match those of the weathered MBAs. The values of the average S-, Sq- and Q-types (DeMeo et al. 2009) were plotted for comparison. The marked range represent a one sigma spread around the average.
\label{fig:Fig12}}
\end{figure}

\begin{figure}
\centerline{\includegraphics[width=8.5cm]{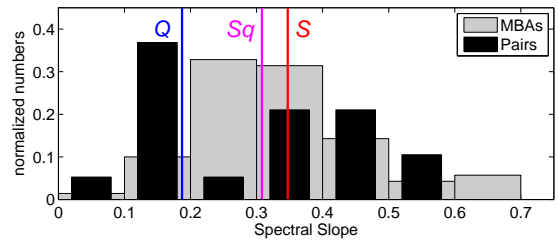}}
\caption{The distribution of the spectral slopes of 19 asteroid pairs (black histograms), 70 MBAs representing the background population (with same size range as the pairs; gray histograms), and the average spectrum of classical groups (red line: S-type, magenta: Sq-type, blue: Q-type; DeMeo et al. 2009). The Kolmogorov-Smirnov test rejects the null hypothesis that the distributions of the spectral slopes of the pairs and the background asteroid population are drawn from the same distribution at $>90\%$ confidence level, supporting the idea that asteroid pairs as a group are unique as far as weathering effects are concerned.
\label{fig:Fig13}}
\end{figure}

	We compare the slope-depth values of the pairs to those of 48 OC meteorites (Dunn et al. 2010), 178 NEAs (from the SMASS survey; Binzel et al. 2004) and 70 main belt asteroids in the same size range of the pairs\footnote{We limit the background population of MBAs to be in the same size of the pairs since it is known that asteroid size is correlated with some spectral parameters (band depth and slope, for example; Gaffey et al. 1993). We distinguish the NEAs from the MBAs since the surfaces of NEAs are modified by physical processes that are irrelevant for MBAs (planetary encounters; Binzel et al. 2010).} ($1<D<15$ km; DeMeo et al. 2009). This comparison shows that 54827 and 8306 have meteoritic-like slopes and therefore they likely present fresh surfaces; 19289, 44612, 88604, 3749, 60546 and 1741 have relatively fresh spectral slopes that can be found on NEAs of the Q-type taxonomy and are rare at the main belt; the other 11 pairs in our sample have spectral slopes that are typical for MBAs (Fig.~\ref{fig:Fig12}). The Kolmogorov-Smirnov test rejects the null hypothesis that the pairs’ distribution of spectral slopes is drawn from the same distribution of the background asteroid population in the main belt at $>90\%$ confidence level, supporting the idea that asteroid pairs as a group are unique as far as weathering effects are concerned (Fig.~\ref{fig:Fig13}). The distribution of the pairs’ band depth is not distinct compared to the background population, most probably because it is not a function of the weathering state alone, as mentioned above.

\subsection{Correlating Spectral Slopes with Age}
\label{sec:slopesAge}

\begin{figure}
\centerline{\includegraphics[width=8.5cm]{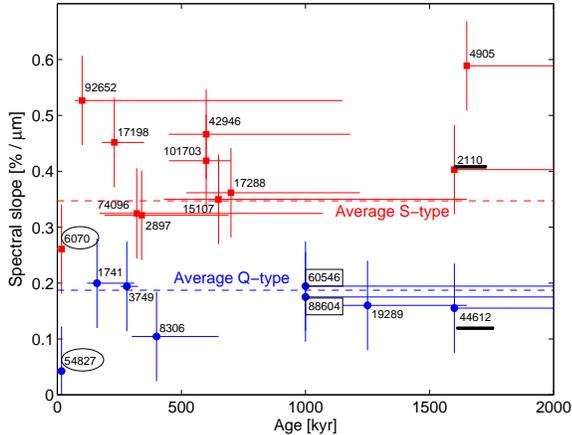}}
\caption{The spectral slopes vs. the ages of the pairs. Asteroids with spectral slope lower than 0.2 are marked in blue circles while those with spectral slopes higher than 0.2 are marked in red squares. The spectral slopes of the average Q-type and S-type are marked with blue and red dashed-lines, respectively. The primary and secondary which belong to the same pair are marked (2110-44612 - underline; 6070-54827 - ellipse; 88604-60546 - rectangle). No correlation is noticeable. However, lower limits on the timescale of space weathering can be derived: the secondary 54827 has a very low slope that put a lower limit of a few $10^4$ years on the time space weathering starts being effective (see text). In addition, pairs with Q-type-like, relatively fresh slopes have a wide range of ages within the entire checked period (of 2 million years). Finding objects that retain their fresh surfaces during this time range provides an indication that space weathering can be as ``slow" as 2 million years for the time it takes an asteroid to present a weathered surface, with slope higher than $0.2\%$ per $\mu m$, as seen on S-type MBAs.
\label{fig:Fig14}}
\end{figure}

	In order to estimate the timescale of space weathering, we search for correlation between the spectral slopes and the ages of the pairs (Fig.~\ref{fig:Fig14}). No clear correlation is present: low and high spectral slopes exist for both young and old asteroid pairs. Furthermore, primaries and secondaries of a single pair, that obviously have the same age, present significant differences between their spectral slopes. The following section discusses this result.

\section{Discussion}
\label{sec:discussion}

	Here we discuss several alternative interpretations and possible ways to understand our results.

\subsection{Observational Variance}
\label{sec:problemsObservations}

	No observational measurements are perfect. Variations or even errors can be caused by using a wrong standard star or observing in bad weather. These can have large-scale effects on the resulting reflectance spectrum and especially on the spectral slope. However, most asteroids reported here were calibrated using multiple standard stars, and some were observed on different nights (Table~\ref{tab:ObsCircum}), reproducing the same results within the uncertainty range. In addition, Wolters (private communication) also observed several of the asteroids reported here and derived the same taxonomic interpretations. As a precaution we estimated the systematic error using the standard deviation of the spectra of 2110 and 3749 that were measured on different nights, although the uncertainty dictated by the SNR of each spectrum is much smaller. Therefore, the reflectance spectra published here are most probably correct within a few percent.

	{\it Phase reddening} (Sanchez et al. 2012; Reddy et al. 2012b) can slightly modify the spectral parameters of the asteroids. However, more than $80\%$ of our observations were obtained at phase angles lower than 15 degrees and the maximal observed phase angle was 37 degrees. Over the narrow range, phase reddening effects are negligible.

	The surface temperature of the asteroid can modify the band center (e.g., Sanchez et al. 2012, Dunn et al. 2013). However, since all of the asteroids in our sample reside in the cold main-belt, any correction to the band center will affect all of them in the same manner (however, this is a good reason not to compare band centers of the hotter NEAs to those of the colder MBAs without proper calibration).

	A possible source of concern is the use of taxonomical archetypes for the visible section of the spectrum. To overcome this uncertainty we chose a conservative approach and used the two possible extremes of weathered and non-weathered.

	In some of the cases, the uncertainty in the age estimation of the pairs is large. This can also affect statistical interpretation of the results. However, given the number of pairs in our sample it is unlikely that our general conclusions would be entirely mistaken.

	Finally, one can suggest that the studied asteroids are not dynamically connected and should not have been defined as asteroid pairs. However, since {\it i)} the method used to define these asteroids is backed by convergence of the primaries and secondaries back in time using dynamical methods that were tested on many independent studies, {\it ii)} the rotation periods of the studied asteroids are correlated to the mass ratio between the secondaries and primaries in a way that match the rotational-fission model, and {\it iii)} the distribution of the pairs’ spectral slope is different than the distribution of the background population (Fig.~\ref{fig:Fig13}), we are confident in the ``pairs" nature of these asteroids.

\subsection{Constraints by the space weathering mechanism}
\label{sec:problemsSpaceWeathering}

	Space weathering is a code name for several factors that affect different materials in different ways and rates, and result in different modification levels of a reflectance spectrum (Clark et al. 2002, Chapman 2004, Brunetto et al. 2006, Gaffey 2010). Several common guidelines can be used when dealing with this mechanism. One of them is the dependency on the mineralogy of the weathered surface.

	The ordinary chondrite classification includes a variety of sub-classes: it is known from laboratory experiments that material rich with olivine (compared to orthopyroxene) is more prone to become weathered and at faster rates (Sasaki et al. 2002; Marchi et al. 2005). Vernazza et al. (2009) have modeled the olivine ratio (ol/[ol+opx]) of a sample of asteroids belonging to dynamically young asteroid families and showed a linear relation between the spectral slope and the olivine ratio, where higher spectral slopes are correlated with higher olivine ratio. Following this, we chose from our sample of asteroid pairs only those that are rich with olivine, assuming that only the spectral slopes of these asteroids have any meaning in the space-weathering context.

	The amount of olivine on the asteroid surface can be determined by the band center: The band center of olivine around ~1 $\mu m$ is shifted to higher values compared to those of orthopyroxene (e.g., Gaffey et al. 1993; Dunn et al. 2013; Sanchez et al. 2014). Therefore, we used the band center after removing the continuum (i.e., after de-weathering) as a representative of the olivine ratio. However, normalizing the spectral slope by the continuum-removed band center does not show any correlation with the ages of the pairs. Moreover, the pairs with relatively small band center values (1741, 4905, 42946, 17198, 88604, 101703), that are supposedly less affected by space weathering, present a wide range of spectral slopes, of weathered and fresh surfaces alike. Therefore, we reject the notion that differences in the amount of olivine on the pairs of our sample explain why their spectral slopes do not correlate with their ages.

	Since different types of weathering mechanisms exist (such as the case of (433) Eros, that presents darker craters with no color alteration; Gaffey 2010) one might argue that the weathered pairs are not really weathered rather their spectral parameters reflect other parameters (e.g., grain size). Such a notion might not hold considering the differences in spectral slopes between primaries and their secondaries in our sample. Since a primary and a secondary are most likely have the same composition then if the primaries do not react to “color-changing” space weathering effects than their secondaries should not react to it as well and should have the same slopes.

	Since one of the agents of space weathering are solar wind particles the heliocentric distance might be another factor relevant for the spectral slopes of the pairs. However, the range of the semi major axis of the observed pairs is limited to the main-belt between 2.1 to 2.9 AU with no correlation to the spectral slopes.

	Another possibility is that the timescale of space weathering is significantly shorter than the ages of the pairs. However, since our sample does not present only high spectral slopes we cannot state that all of these pairs are weathered, indicating a fast space weathering. The pair 6070-54827 is very young, most probably separated from each other $\sim17$ thousands years ago and indeed the secondary 54827 has a very low slope ($0.04\%\pm0.08$ per $\mu m$) that matches those found on meteorites and NEAs (we deal with 6070 later on). This puts a lower limit of a few times $10^4$ years on the time space weathering starts being effective. This limit is within the lower limit of $10^5$ years that was suggested by Nesvorn{\'y} et al. (2010) as the time a fresh asteroid will maintain its non-weathered appearance, thus our observation timescale resides within their conclusion. On top of that, pairs with Q-type-like, relatively fresh slopes ($<0.2\%$ per $\mu m$) have a wide range of ages within the entire checked period (of 2 million years). Finding objects that retain their fresh surfaces during this time range provides an indication that space weathering can be as ``slow" as 2 million years for the time it takes an asteroid to present a weathered surface, with slope higher than $0.2\%$ per $\mu m$, as seen on S-type MBAs. Therefore, an extremely fast weathering process is not the source of the weathered pairs in our sample.

	A fourth scenario deals with a possible ``saturation" of the weathered asteroid, meaning that the sub-surface material of the asteroid is also weathered - exposing it will not present any signature of fresh spectrum. If asteroid pairs are subjected to many rotational-fission events during their lifetime, we can assume that each event was followed by a ``gardening" process that lifted dust from the surface that later re-accumulated back on the primary asteroid. After each event the surface becomes weathered until no fresh material can be found on the outside layers of the asteroid. When new fission occurs the raising and re-settling dust that follow it are already weathered, and no fresh material can be observed. However, since the secondary members are of order hundreds of meters to a few kilometers in size, it is hard to imagine that material so deep inside of the progenitor body was ever exposed to the agents of space weathering. Laboratory measurements of grains from 25143 Itokawa (Noguchi et al. 2011) and of lunar samples (summarized in Clark et al. 2002) show that the depth of the weathered layer is on the order of tens to hundreds nanometers; forming this weathered coating on every grain in an approximate volume of a cubic kilometer seems unlikely.

	Since space weathering issues of timescale, saturation and relevant parameters cannot explain the existence of weathered and fresh asteroid pairs, we now discuss the implications of this result on our understanding of the rotational-fission mechanism.

\subsection{Constraints by rotational-fission models}
\label{sec:problemsRotationalFission}

	We, of course, cannot be sure that a rotational-fission event is necessarily followed by the exposure of fresh material on the surfaces of the pairs' members. Thus we discuss three alternative scenarios to explain the existence of weathered asteroid pairs: a gradual fission; a mixture of weathered and fresh material; and the lack of a secondary fission event.

	\textit{\textbf{Gradual fission:}} Walsh et al. (2008) suggested a gradual fission model where rocks and boulders on the surface of the fast-rotating asteroid are shifting and rolling towards the equator, forming an equatorial ridge, before being ejected. While in orbit they accumulate into a satellite that might escape to form an asteroid pair. Since this mechanism is limited to the surface of the asteroid no deep sub-surface material is being exposed. The shifted rocks are relatively small; therefore the extent of fresh material exposed might be small as well and the amount of released dust will be minimal. In addition, since the fission in this model is a long process, a fresh-exposed surface might become weathered before a consecutive fission event will expose another area of sub-surface material. Asteroid pairs that were formed by the gradual model might not present fresh material at all.

	\textit{\textbf{Mixture:}} In the model of a single detached secondary (Scheeres 2007), as the progenitor split and fresh material is ejected and re-accumulates on the primary surface, its previous weathered coating is probably not covered completely by the accumulating fresh dust. Therefore, the asteroid's surface could contain a mixture of fresh and weathered material that displays an intermediate spectrum between fresh and weathered spectrum. If the amount of resettled fresh dust is larger the reflectance spectrum will look fresher and vice versa.

	\textit{\textbf{Secondary fission:}} Jacobson \& Scheeres (2011) suggested that after the detachment of the secondary, it orbits the primary in an unstable orbit until it is lost and the system becomes an asteroid pair. During its short life as a satellite, the primary's tidal forces deform the secondary until it breaks apart by itself into two or more components. This secondary fission includes the exposure of additional fresh dust that re-settles on the primary and recoats it with fresh material. However, in a case of a ``fast fission", the secondary will be lost before it breaks apart, and the amount of fresh dust will be decreased significantly. Jacobson \& Scheeres (2011) found that the time elapsed after the detachment of the secondary until it completely leaves the vicinity of its primary has a normal distribution with extremes of few days to tens or a few hundreds of days (see their Figs. 7 and 8). In this scenario, the pairs observed by our study to have high spectral slopes might have lost their secondaries quickly, avoiding the secondary fission phase, and thus maintained a higher percentage of their original weathered surfaces. 

\begin{figure}
\centerline{\includegraphics[width=8.5cm]{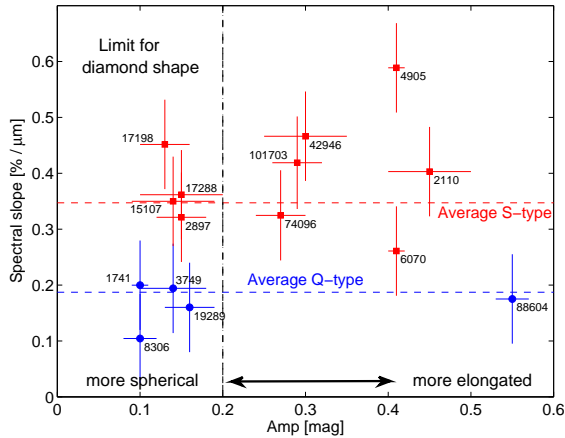}}
\caption{The spectral slopes vs. the lightcurve amplitude of 15 primary pairs from our sample (the amplitude of 92652 is unknown). The gradual fission model (Walsh et al. 2008) produces ``diamond-shape" primaries with almost spherical shapes, that have small amplitude on their lightcurves (in the order of $<0.1-0.2$ mag). Therefore, only four asteroids among the weathered primary pairs in our sample (2897, 15107, 17198, 17288) can be explained by the gradual model, while the fast model can explain them all.
\label{fig:Fig15}}
\end{figure}

	Can any of these three scenarios be supported or rejected by our observations? The \textit{\textbf{gradual model}} of Walsh et al. produces mostly spherical objects such as (66391) 1999 KW4, known for its diamond shape (Ostro et al. 2007, Harris et al. 2009). These objects present little variation in their brightness while they spin, therefore, the amplitude of their lightcurve is small\footnote{Amplitude of well-known ``diamond-shape" asteroids: 0.12 mag for (66391) 1999 KW4 is (Pravec et al. 2006); 0.17 mag for (101955) Bennu (Hergenrother et al. 2012).}. Asteroid pairs that disintegrate in a single, fast fission event could result in elongated or spherical objects alike, even though, elongated asteroids are more prone to break apart by rotational fission and are more probable to maintain some of their elongation after the fission as well (Pravec et al. 2010). Therefore, these asteroids can present lightcurves with high amplitude. The amplitude values of 15 out of the 16 OC primary components in our sample were published (Polishook 2011, Polishook et al. 2011, Polishook 2014, Pravec et al. 2010, Slivan et al. 2008, Warner et al. 2009, and Pravec’s web site: http://www.asu.cas.cz/$\sim$ppravec/) and are presented on Table~\ref{tab:PairsParam}. Plotting the spectral slopes as a function of the amplitude of the pairs' lightcurves (Fig.~\ref{fig:Fig15}) shows that while four asteroids (2897, 15107, 17198, 17288) have high spectral slopes and low amplitude, six other pairs (2110, 4905, 6070, 42946, 74096, 101703) with high slopes have amplitude values that are much higher than expected for the diamond-shape asteroids form by Walsh et al. model. The gradual model could not explain the primary pairs with low spectral slopes either. Therefore, the gradual fission model discussed here might explain only a quarter of the observed pairs. 

\begin{figure}
\centerline{\includegraphics[width=8.5cm]{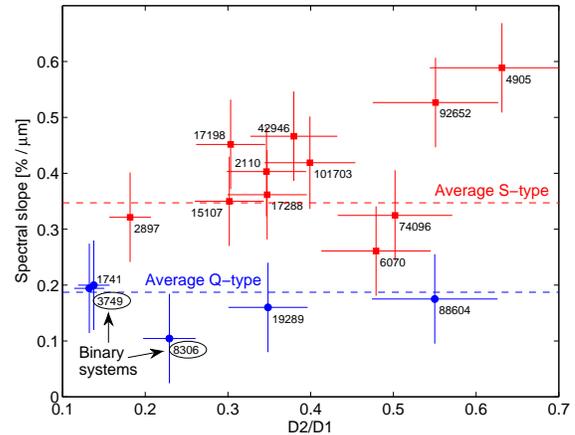}}
\caption{The spectral slopes vs. the size ratio (D2/D1) of the 16 primary pairs in our sample. Primary pairs with low spectral slope were covered by a significant amount of fresh material. If a single fission occurs than pairs with high D2/D1 should present low spectral slope (such as 88604); if the secondary member disintegrate due to tidal forces from the primary, the pairs with low D2/D1 stands for further disrupted secondaries and should present low spectral slope as well (such as 1741 and 3749). Therefore, the lack of correlation between the spectral slope and the pairs size ratio, do not give a conclusive result in favor of one of the models. However, Jacobson \& Scheeres (2011) have predicted that $\sim40\%$ of the pairs should run through a secondary fission; if low spectral slope is a marker for secondary fission, than $33\%$ of the pairs in our sample had their secondaries disintegrate, the same order of magnitude as predicted by Jacobson \& Scheeres. Jacobson \& Scheeres also predict that a secondary fission can result with the formation of a binary asteroid. Indeed, two of the primaries with low D2/D1 values and low spectral slopes have known satellites (3749 and 8306; marked by ellipses), supporting the secondary fission model.
\label{fig:Fig16}}
\end{figure}

	\textit{\textbf{Mixing}} fresh and weathered dust may explain why the primaries in our sample present weathered spectral slopes compared to their fresh-looking secondaries. Both 2110 and 6070, have slopes of $0.40\pm0.08$ and $0.26\pm0.08$, respectively, while their secondaries, 44612 and 54827, present low spectral slopes of $0.16\pm0.08$ and $0.04\pm0.08$, respectively. Moreover, this explains why the young ($17x10^4$ years) asteroid 6070 does not show a nearly zero slope. Since the amount of re-accumulating fresh dust is a special case for each asteroid, some of the primaries might have reflectance spectra with higher spectral slopes than 6070, or very low slopes such as 54827. 88604-60546 is such a pair where the two members have the same low slope within the uncertainty range. This suggests that the primary was covered by a significant amount of fresh material. The fact that the secondary of this pair is large (D2 = 3.4 km, D2/D1 = 0.55) supports the notion that a significant amount of dust was released during the fission of this pair’s progenitor. However, the other pairs in our sample (4905, 6070, 74096, 92652) with high size ratio ($D2/D1>0.4$) present high spectral slopes that are harder to explain with the mixing scenario (Fig.~\ref{fig:Fig16}).

	Alternatively, if the \textit{\textbf{secondary fission}} is the significant source of the fresh dust, as suggested by the third model we examine, then smaller D2/D1 is the result of a ``slow" fission that continuously disrupts the secondary and releases higher amounts of fresh dust that re-coats the primary with material of lower spectral slope. This can explain the low spectral slopes seen on primary pairs with small D2/D1 (such as 1741, 3749 and 8306), and the high spectral slopes seen on primary pairs with high D2/D1 (such as 4905, 6070, 74096, 92652) even though not all pairs follow this rule (such as 88604).

	It is interesting to note that according the model of Jacobson \& Scheeres (2011) $40\pm4\%$ of pairs with low mass ratio ($M2/M1<0.2$, i.e. 15 out of 16 pairs in our sample, excluding 4905) will undergo a secondary fission. If indeed pairs that undergo a secondary fission are marked by a low spectral slope then $33\%$ of the pairs in our sample had their secondary disintegrated, very similar to the prediction of Jacobson \& Scheeres. Moreover, disrupted secondaries can form a second satellite around the primary and Jacobson \& Scheeres describe how these systems could maintain one of the satellites when the other escapes, defining the system as a binary and a pair system. Indeed, among the five primary pairs in our sample with low spectral slope, two pairs have known satellites - 3749 (Merline et al. 2002b, Marchis et al. 2008) and 8306 (Polishook 2014; Pravec private communication). Furthermore, the primary of each asteroid has a low D2/D1, suggesting that their secondaries were continuously fissioned. Therefore, our observations provide the most support for the Jacobson \& Scheeres (2011) \textit{\textbf{secondary fission}} model.

\subsection{Constraints on the space weathering timescale}
\label{sec:SWtimescale}

	If the spectral slope of the primary member of the pair is a function of the amount of released fresh dust that resettled on the original weathered surface, and not of the age of the pair, then the timescale of space weathering cannot be derived until the fission parameters (sizes, shapes, etc.) of both members is measured. However, a first-order fit could be derived by examining the secondaries only. Consisting from sub-surface material of the progenitor body, these objects should have fresh surfaces with lower spectral slope and due to their smaller sizes these bodies do not attract additional dust as the primary members, hence they serve as ``un-corrupted" samples. Unfortunately, our current sample includes only three secondaries, and only lower limits are known for two of them, therefore, our estimation of space weathering timescale is not well constrained. Even though, as described above, within our sample we can define a minimal limit of 2 million years for the time an object with fresh surface will present high spectral slope that is typical for S-type asteroids. By focusing on the spectral parameters of secondary pairs, future studies might derive more generally the timescale of the space weathering mechanism.

\section{Conclusions}
\label{sec:conclusions}

	An asteroid pair consists of two unbound components which separated in the last $\sim1-2$ Myrs from a single ``rubble-pile" structured progenitor, that fails to remain bond against a fast rotation. Models suggest that this rotational-fission process likely involves the exposure of material from below the progenitor surface - these materials may have never been exposed to the weathering conditions of space and therefore at least initially present non-weathered, fresh spectra.

	We have measured the near-infrared spectra of 31 asteroids in pairs, collected their visible spectra or broadband photometric colors, and analyzed them to derive their spectral slopes and band parameters. These values were used to estimate the weathering state of these asteroids. In addition, we used dynamical calculations to estimate the age of the pair, namely the time that passed since the fission of the progenitor.

	Our measurements show that 19 of the pairs in our sample are S-complex (OC, Ordinary Chondrites), 6 are C/X-complex, and 6 are V-type asteroids. The variety of taxonomic types shows that the composition of the asteroid is irrelevant for the rotational-fission mechanism that is effective for asteroids with a ``rubble-pile" structure.

	In the four cases where we observed both the primary and the secondary members of a pair, both presented the spectra of the same taxonomy. This is consistent with a common origin of components in each of these asteroid pairs.

	The two Q-type asteroids in our sample (19289 and 54827) are the first of their kind to be observed in the main-belt of asteroids over the full visible and near-infrared spectral range. This solidly demonstrates that the Q-type taxonomy is not limited to the NEA population.

	Eight pairs out of the nineteen observed OC asteroids present low spectral slopes of less than $0.2\%$ per $\mu m$, the maximal limit for the slope of meteorites and approximately the spectral slope of an average Q-type. This supports the notion that the rotational-fission mechanism can be involved with the exposure of fresh, sub-surface material.

	There is no clear evidence for a correlation between the spectral parameters and the ages of the pairs. However, our sample includes “old” pairs ($2x10^6 \ge $age$ \ge 1x10^6$ years) that present relatively low, Q-type-like spectral slopes ($<0.2\%$ per $\mu m$). This illustrates for these asteroids a timescale of at least $\sim2$ million years to develop high spectral slope that is typical for S-type asteroids.

	We describe three alternative scenarios of the rotational-fission mechanism that explain why 11 of the OC pairs in our sample present high slopes that indicate weathered surfaces:

{\it 1.}	A gradual rotational-fission includes the ejection of rocks and boulders from the surface that re-accumulate into a secondary pair while in orbit around the primary. Since most of the transportation of material is on the surface, and since this kind of fission can take $\sim10^6$ years to conclude, there is no significant amount of fresh dust released. In addition, this model results with almost spherical asteroids, recognized by their low amplitude on their lightcurves. However, only four of the pairs in our sample with high slopes have low lightcurve amplitude, therefore, only a fraction of our sample could be explained by this model.

{\it 2.}	A mixture between the ejected fresh dust and the original weathered surface presents a superposition of the two spectra depending on the ratio between the fresh and weathered material. This explains why the primaries in our sample have higher spectral slope than the secondaries. However, while it is expected that pairs with high D2/D1 present low spectral slope, most of the pairs in our sample do not follow this rule.

{\it 3.}	A continuous fission of the secondary while it orbits the primary subsequent to the original fission (Jacobson \& Scheeres 2011), releases even more fresh material compared to fast fission events where this phase is skipped. Therefore, high size ratio D2/D1 probably means that the secondary did not disintegrate and less amount of fresh dust is released. Indeed, most, though not all, of the pairs in our sample follow this rule. In addition, $33\%$ of the pairs in our sample have low spectral slope - this is similar to the prediction of Jacobson \& Scheeres of $40\%$ of the pairs that run through a secondary fission. Two out of the five pairs in our sample with low spectral slope have known satellites on top of being a pair, which further confirms the theoretical model of secondary fission.

\section{Open Questions and Future Work}
\label{sec:futureWork}

	Open questions remain and new questions arise on the physics of asteroid pairs:

\begin{itemize}
\item	Did all pairs form by the fast fission model, or were some formed by the gradual process? Why do some of the pairs evolve in one way and not the other? What are the relevant parameters for the fission of the secondary object to take place? What could this tell us about the internal structure of asteroids?
\item	If pair primaries with low spectral slopes are markers for a binary system, satellites should be searched for around them. A photometric survey to detect mutual events within the lightcurves of primary pairs could give solid support for the model of secondary fission and could inform the ratio of binary asteroids among asteroid pairs and vice versa.
\item	What is the amount of fresh material needed to ``paint" a weathered surface so it will have low spectral slope? By modeling specific systems of pairs we could better understand the physical context of fresh and weathered asteroidal surfaces.
\item	If the secondary components in pairs are indeed excavated from sub-surface material of the progenitor body, could measuring their spectral slopes reveal the timescale of space weathering?
\item	If the immediate-splitting model is valid, young primary pairs might have hemispherical color asymmetries. The area of fresh separation might have a lower spectral slope compared to other areas on the asteroid. If the gradual model is more common, then the areas of exposed fresh material will be too small to detect, and no area with low spectral slope will be observed. Therefore, a rotational resolved spectral study of young asteroid pairs might give additional support to either one of the formation models discussed above. We have started a campaign of rotational resolved spectroscopy of asteroid pairs and will present our results in a separate paper.
\item	A question that cannot be addressed by spectral observations alone is what is the frequency of asteroid pairs among the entire population of asteroids. Though challenging, answering this question will reveal the fraction of ‘rubble pile’ asteroids and the way they are formed, evolve and break apart, making this question worth a study of its own.
\end{itemize}

\acknowledgments

	We thank Seth Jacobson, Kevin Walsh, Paul Sanchez and Stephen Wolters for fruitful discussions, Petr Pravec for sharing with us his pre-published data and Brad S. Cenko for his observations at the Lick Observatory. We thank Vishnu Reddy and Mark Willman for their useful reviews.

	DP is grateful to the AXA research fund for their generous postdoctoral fellowship. NM would like to acknowledge support from the National Science Foundation Astronomy and Astrophysics Postdoctoral fellowship program. RPB acknowledges NASA Near-Earth Object Observation program support through grant NNX10AG27G. FED acknowledges support by the National Science Foundation under Grant 0907766 and by NASA under Grant NNX12AL26G and through Hubble Fellowship grant HST-HF-51319.01-A awarded by the Space Telescope Science Institute, which is operated by the Association of Universities for Research in Astronomy, Inc., for NASA, under contract NAS 5-26555. The work of DV and J\v{Z} was supported by the Czech Grant Agency (grant 209/12/0229). DO was supported by Polish National Science Center, grant number NCN 2012/04/S/ST9/00022. This study was supported by the National Science Foundation under Grant 6920422. Any opinions, findings, and conclusions or recommendations expressed in this material are those of the authors and do not necessarily reflect the views of the National Science Foundation.

	Observations reported here were obtained at the Infrared Telescope Facility, which is operated by the University of Hawaii under Cooperative Agreement NCC 5-538 with NASA, Science Mission Directorate, Planetary Astronomy Program. This paper also includes data gathered with the 6.5 meter Magellan Telescopes located at Las Campanas Observatory, Chile. and with the Nordic Optical Telescope, operated by the Nordic Optical Telescope Scientific Association at the Observatorio del Roque de los Muchachos, La Palma, Spain, of the Instituto de Astrofisica de Canarias. We appreciate the Hawaiian, Chilean, Californian, Spanish and Israeli people for allowing us to use their sacred mountains for astronomy.

\end{document}